\documentclass[a4paper,11pt]{article}
\pdfoutput=1 

\usepackage{jcappub} 

\usepackage[T1]{fontenc} 
\usepackage{amsmath, amssymb, bm, graphicx, graphics, color, mathrsfs}
\RequirePackage{graphicx}
\RequirePackage{mathptmx}      
\RequirePackage{flushend}
\RequirePackage[numbers,sort&compress]{natbib}

\title{\boldmath Dark matter influence on black objects thermodynamics}

\author[a,1]{Marek Rogatko,\note{Corresponding author.}}
\author[a]{Aneta Wojnar}

\affiliation[a]{Institute of Physics,
Maria Curie-Sk{\l}odowska University \\
20-031 Lublin, pl. Marii Curie-Sk{\l}odowskiej 1, Poland}

\emailAdd{rogat@kft.umcs.lublin.pl}
\emailAdd{aneta.wojnar@poczta.umcs.lublin.pl}

\abstract{Physical process version of the first law of black hole thermodynamics in Einstein-Maxwell {\it {\it dark matter}  }  gravity was derived. The {\it dark matter} sector is mimicked by
the additional U(1)-gauge field coupled to the ordinary Maxwell one. By considering any cross section of the black hole event horizon to the future of the bifurcation surface,
the equilibrium state version of the first law of black hole mechanics was achieved. The considerations were generalized to the case of Einstein-Yang-Mills {\it dark matter} gravity theory.
The main conclusion is that the influence of {\it dark matter} is crucial in the formation process of black objects. This fact may constitute the explanation of 
the recent observations  of the enormous mass of the super luminous quasars formed in a relatively short 
time after Big Bang. We also pay attention to the compact binaries thermodynamics, when {\it dark matter}
sector enters the game.}

\begin{document}
\maketitle
\flushbottom
\newcommand  {\Rbar} {{\mbox{\rm$\mbox{I}\+!\mbox{R}$}}}
\newcommand  {\Hbar} {{\mbox{\rm$\mbox{I}\!\mbox{H}$}}}
\newcommand {\Cbar}{\mathord{\setlength{\unitlength}{1em}
     \begin{picture}(0.6,0.7)(-0.1,0) \put(-0.1,0){\rm C}
        \thicklines \put(0.2,0.05){\line(0,1){0.55}}\end {picture}}}
\newcommand{\be}{\begin{equation}}
\newcommand{\ee}{\end{equation}}
\newcommand{\ben}{\begin{eqnarray}}
\newcommand{\een}{\end{eqnarray}}

\newcommand{\la}{{\lambda}}
\newcommand{\Om}{{\Omega}}
\newcommand{\ta}{{\tilde a}}
\newcommand{\bg}{{\bar g}}
\newcommand{\bh}{{\bar h}}
\newcommand{\bdel}{{\bar \delta}}
\newcommand{\si}{{\sigma}}
\newcommand{\C}{{\cal C}}
\newcommand{\D}{{\cal D}}
\newcommand{\cA}{{\cal A}}
\newcommand{\cT}{{\cal T}}
\newcommand{\cO}{{\cal O}}
\newcommand{\eeo}{\cO ({1 \over E})}
\newcommand{\G}{{\cal G}}
\newcommand{\cL}{{\cal L}}
\newcommand{\cH}{{\cal H}}
\newcommand{\cE}{{\cal E}}
\newcommand{\cM}{{\cal M}}
\newcommand{\cR}{{\cal R}}
\newcommand{\cB}{{\cal B}}
\newcommand{\cJ}{{\cal J}}

\newcommand{\p}{\partial}
\newcommand{\na}{\nabla}
\newcommand{\ssum}{\sum\limits_{i = 1}^3}
\newcommand{\dssum}{\sum\limits_{i = 1}^2}
\newcommand{\tal}{{\tilde \alpha}}
\newcommand{\ints}{\int_{\Sigma} d\Sigma}
\newcommand{\LieN}{{\cal L}_{N^{i}}}
\newcommand{\Lief}{{\cal L}_{\phi^{i}}}
\newcommand{\Liet}{{\cal L}_{t^{i}}}
\newcommand{\LieM}{{\cal L}_{M^{\mu}}}
\newcommand{\Lie}{{\cal L}}

\newcommand{\tpe}{{\tilde p}}
\newcommand{\tp}{{\tilde \phi}}
\newcommand{\tPhi}{\tilde \Phi}
\newcommand{\tpsi}{\tilde \psi}
\newcommand{\tchi}{\tilde \chi}
\newcommand{\tim}{{\tilde \mu}}
\newcommand{\tom}{{\tilde \omega}}
\newcommand{\tr}{{\tilde \rho}}
\newcommand{\tV}{{\tilde V}}
\newcommand{\tir}{{\tilde r}}
\newcommand{\rp}{r_{+}}
\newcommand{\hr}{{\hat r}}
\newcommand{\rv}{{r_{v}}}
\newcommand{\dr}{{d \over d \hr}}
\newcommand{\dR}{{d \over d R}}

\newcommand{\hhf}{{\hat \phi}}
\newcommand{\hhM}{{\hat M}}
\newcommand{\hhQ}{{\hat Q}}
\newcommand{\hht}{{\hat t}}
\newcommand{\hhr}{{\hat r}}
\newcommand{\hhS}{{\hat \Sigma}}
\newcommand{\hhD}{{\hat \Delta}}
\newcommand{\hhm}{{\hat \mu}}
\newcommand{\hro}{{\hat \rho}}
\newcommand{\hhz}{{\hat z}}

\newcommand{\hI}{\hat I}
\newcommand{\hg}{\hat g}
\newcommand{\hR}{\hat R}
\newcommand{\hD}{\hat D}
\newcommand{\hna}{\hat \nabla}
\newcommand{\tF}{\tilde F}
\newcommand{\tT}{\tilde T}
\newcommand{\tL}{\tilde L}
\newcommand{\hC}{\hat C}

\newcommand{\ep}{\epsilon}
\newcommand{\tep}{\tilde \epsilon}
\newcommand{\bep}{\bar \epsilon}
\newcommand{\ppp}{\varphi}
\newcommand{\Ga}{\Gamma}
\newcommand{\ga}{\gamma}
\newcommand{\hth}{\hat \theta}
\newcommand{\zpsi}{\psi^{\ast}}

\newcommand{\Dsl}{{\slash \negthinspace \negthinspace \negthinspace \negthinspace  D}}
\newcommand{\tD}{{\tilde D}}
\newcommand{\tB}{{\tilde B}}
\newcommand{\talpha}{{\tilde \alpha}}
\newcommand{\tbeta}{{\tilde \beta}}
\newcommand{\hT}{\hat T}
\section{Introduction}
\label{sec:introduction}

In our paper we shall pay attention to the problem of the first law of thermodynamics for Einstein-Maxwell {\it dark matter} gravity, where the {\it dark matter} sector will be 
mimicked by the $U(1)$-gauge field coupled to the Maxwell one. The main motivation standing 
behind our research is to explain the possibility of existence of supermassive black object in the early Universe which is the puzzle for the contemporary high energy
astrophysics. The tantalizing question is related to the debate how such a supermassive object can be created in a relatively short time after Big Bang. Perhaps accretion of 
{\it dark matter} by black objects might explain these facts.

Black hole thermodynamics constitutes one of the key subject of the mathematical theory of black holes, both in four-dimensional Einstein gravity like in its
generalizations. The subject in question is referred \cite{wal94} as two versions. The first one, the so-called {\it
equilibrium state version}  and the other, {\it physical process} one. The {\it equilibrium state version} formulation was given 
in the seminal paper of Bardeen, Carter and Hawking \cite{bar73}, and it is 
with the linear
perturbations of a stationary electrovacuum black hole to another stationary black hole state.
Arbitrary asymptotically flat perturbations of a stationary black hole were elaborated in \cite{sud92}. On the other hand, 
an arbitrary diffeomorphism invariant Lagrangian with a metric and matter fields being subject to stationary and axisymmetric
black hole solutions, was treated in \cite{wal93}-\cite{iye97}. The higher curvature and higher derivative terms of the aforementioned
problem were given in \cite{jac94, jac95}, while the case of the Lagrangian being an arbitrary function of a metric,
Ricci tensor and a scalar field was considered in \cite{kog98}.
The general case of a charged and rotating black hole with matter fields which were not smooth through the event horizon was given in \cite{gao03}.

The first law of black hole thermodynamics was also intensively studied in the case of $n$-dimensional
black holes.  Namely, the {\it equilibrium state} version was elaborated under the assumption of spherical topology of black holes and the supposition that the four-dimensional black 
hole uniqueness theorem extends to a higher dimensional case \cite{tow01}-\cite{rog05a}.

The {\it physical process} version of the first law of thermodynamics for a black object is realized
by changing the stationary black hole state by an infinitesimal physical process, e.g.,
throwing some portion of matter into it. It was supposed that the final state will settle down to a stationary one, and we can extract some information
of the changes of black hole parameters and these facts in turn help us to find the form of the first law of thermodynamics. The {\it physical version} of black hole thermodynamics was studied in 
the realm of Einstein and Einstein-Maxwell gravity in \cite{wal94} and \cite{gao01}. 
The regularity of the solution is the key feature which should be satisfied in these covariant space methods \cite{bar03,bar08}. 

The first law of mechanics for the low-energy limit of string theory black holes (the so-called Einstein-Maxwell-axion-dilaton gravity)
were examined in \cite{rog02}, while the $n$-dimensional black hole \cite{rog05} and black ring, as well as, black Saturn and p-branes first law of mechanics
were elaborated in \cite{cop06}-\cite{rog09}.

Recently the attitude to the first law of black hole mechanics for fields with internal gauge freedom was performed \cite{pra17}. There are some attempts to study
the mechanical properties of accelerating black holes \cite{dut06}-\cite{ast17}.

The thermodynamics of compact binary objects has been also intensively studied. In \cite{Fus} the authors considered black holes' system as well as the one including neutron
stars modeled by perfect-fluid while the magnetized case was examined in \cite{Eric}. Both cases require the notion of helically symmetric spacetimes 
\cite{blackburn, detweiler, bona, asada, Fus, klein, torre1, torre2, beig, bicak, yoshida, whelan} since they approximate the early stage of a binary system. 
Moreover, studies on the magnetized systems demand extra conservation laws applied to the ideal magnetohydrodynamic fluid \cite{bek, bek2} which was also taken into account 
in \cite{Eric}. The stability of binary objects orbits was examined for example in \cite{glenz, wilson1, wilson2, baum}.
Because of the last discoveries of gravitational waves coming from mergers of black holes \cite{gw1,gw2,gw3,gw4,gw6} and 
neutron stars' coalescence \cite{gw5}, the studies on compact binaries are on the high demand.

The organization of our paper is as follows. In section 2 we describe the main features 
of the model under inspection with the two $U(1)$-gauge fields coupled together, paying attention to its possible experimental confirmation, as well as,
string/M-theory justification of the theory in question.  Section 3 is devoted to the equilibrium state version of the first law of thermodynamics for black holes in the considered theory,
whereas in section 4, we elaborate its Yang-Mills generalization. Furthermore, we examine 
the influence of {\it dark matter} sector on the 
thermodynamics of compact binary systems. In section 6 we conclude our investigations.


\section{Influence of the {\it dark matter} on physical process version of first law of thermodynamics}
\subsection{Dark matter model}

The motivation standing behind our research is to elaborate the imprint of the {\it dark matter}  on physical phenomena being
one of the most intriguing question of the contemporary physics. 
The non-baryonic {\it dark matter} comprises over 23 percent of the mass of the observed Universe. {\it Dark matter} builds a thread-like structure
of the cosmic web constituting a scaffolding for the ordinary matter to accrete \cite{mas07,die12}. The first direct glimpse of the shape of the scaffolding was delivered by
Space Telescope Hubble, studying a giant filament of {\it dark matter}, being the part of the comic web, which  extends from one of the most massive galaxy clusters MACS J071 \cite{die12}.

The key prediction of the current understanding of the creation
of structures in the Universe is the so-called $\Lambda$CDM model,
foresees that galaxies are embedded
in very extended massive halos composed of {\it dark matter}, which in turns are surrounded by smaller {\it dark matter} sub-halos. The sub-halo {\it dark matter}
clumps are large enough to accumulate gas and dust in order to form satellite galaxies, which can orbit around the host ones. In principle smaller galaxies can be circled 
by much smaller sub-halo {\it dark matter} satellites, almost invisible to telescopes \cite{sta16}. It leads to the conclusion that in the nearby of the Milky Way one can suppose 
that such kind of structures can also exist.
On the other hand, 
{\it dark matter} interaction with the Standard Model particles
is one of the main theoretical searches of the particle physics in the early Universe \cite{reg15,ali15}. 

Collapse of neutron stars 
and emergence of the first star generations can deliver some other hints for these researches in question \cite{bra14}-\cite{lop14}. The existence of {\it dark matter} can affect
black hole growth during the early stages of our Universe. The numerical studies of {\it dark matter} and dark energy collapse and their interactions with black holes and wormholes
were investigated in \cite{nak12,nak15a}.

Physics beyond the Standard Model, implemented for the explanation {\it dark matter} 
non-gravitational interactions, increases the interests in gamma rays 
emissions coming from dwarf galaxies, possible dilaton-like coupling to photons cau\-sed 
by ultra-light {\it dark matter}, as well as, oscillations of the fine structure constant \cite{ger15}-\cite{til15}.
On the other hand, Earth experiments are also used to detect possible low-energy mass of {\it dark matter} sector,
especially in  $e^+~e^-$ colliders \cite{babar14}. BABAR detector set some energy range for dark photon production, i.e.,
$0.02 < m < 10.2 GeV$, but no significant
signal has been observed. The new experiments are planned to cover the energy region $15 \leq m \leq 30 MeV$. Recently, the revision of the constraints on {\it dark photon} with
masses below $100 MeV$ from the observation of supernova 1987A are delivered \cite{cha17}. 

In our research we shall consider the model of {\it dark matter} sector in which the additional $U(1)$-gauge field is coupled to the ordinary Maxwell one. 
The Lagrangian describing Einstein-Maxwell {\it dark matter} gravity yields \cite{vac91,ach00}
\be
\mathcal{L} = {\bf \ep} ~\bigg( R - F_{\mu\nu}~F^{\mu\nu} - B_{\mu\nu}~B^{\mu\nu} - \alpha~F_{\mu\nu}B^{\mu\nu} \bigg),
\label{lagr}
\ee
where by $ {\bf \ep}$ we denote the volume element, $F_{\mu \nu} = 2 \na_{[\mu }A_{\nu]}$ is the ordinary Maxwell field while $B_{\mu \nu} = 2 \na_{[\mu }B_{\nu]}$ 
stands for the auxiliary $U(1)$-gauge field mimicking the {\it {\it dark matter} } sector which is coupled to Maxwell one. The coupling constant is denoted by $\alpha$.
Predicted values of $\alpha$-coupling constant, being
the kinetic mixing parameter between the two $U(1)$-gauge fields, for realistic string compactifications range between $10^{-2}$ and $10^{-16}$ \cite{abe04}-\cite{ban17}.

The idea of {\it dark matter} sector coupled to the Maxwell one, has a strong astrophysical support provided by observations of $511$ eV gamma rays, 
\cite{integral} , experiments detecting the electron positron excess in galaxy
\cite{atic, pamela}, possible explanation of muon anomalous magnetic moment \cite{muon}.
It was suggested in \cite{dav12} that kinetic mixing term between ordinary boson and relatively light one (the {\it dark} one) arising from $U(1)$-gauge symmetry connected with a hidden sector,
may lead to the new source of low energy parity violation. This phenomenon may be explored by the future atomic parity violation and planned polarized electron scattering experiments.\\
On the other hand, it was claimed in \cite{dav13} that the low energy gauge interaction in the hidden sector may manifest itself by the Higgs boson $H$ decays, i.e.,
the Abelian symmetry breaks, which causes that a relatively light vector boson $Z_d$ with mass $m\ge 10GeV$ can arise. It was assumed that it did not couple to the Standard Model (SM) states. However,
the possibilities of extension of the SM were taken into account, namely a second Higgs doublet, or new heavy leptons carrying {\it dark charge}.  These assumptions leads to the properties of Higgs 
closely to those from SM. The decays of the following types were studied: ~$H \rightarrow X~Z_{dark}$, where $X$ stands for ordinary vector boson $Z$, ~$Z_{dark}$, or gamma quanta.\\
Collisions among galaxy clusters can also provide new tools for testing non-gravitational forces 
acting on {\it dark matter} \cite{massey15a}.

On the other hand, the model in question has its justification in string/M-theory, where the mixing portal (term which couples Maxwell and the additional $U(1)$-gauge field) arises
in open string theory, where both gauge states are supported by D-branes separated in extra dimensions \cite{ach16}. It takes place, e.g., in supersymmetric Type I, Type II A, Type II B models.
The massive open strings stretch between two D-branes. The massive string/brane states existence connect the different gauge sectors. Another realization of the above scenario can be performed by 
M2-branes wrapped on surfaces which intersect two distinct codimension four orbifolds singularities. The construction in question has its natural generalization in M, F-theory and heterotic string models.

\subsection{Physical version of first law}
The main motivation for our research will be the problem of the appearance of supermassive black holes at very early stages of the Universe history. The recent observations reveal
almost forty quasars at the distances greater than the redshift equal to six. Their masses are in the range of 12 to 17 billion solar masses \cite{fan13}-\cite{tre11}. 
The question posed by the observations is how such huge black objects can be formed in 
a relatively short time after Big Bang.

We conjecture that studies of the first law of black hole thermodynamics in the theory of {\it dark matter} with the additional gauge field coupled to the ordinary Maxwell one can support 
the possible answer to this problem.

In this subsection we shall find mathematically rigorous mass formula for black objects influenced by {\it dark matter}, i.e.,
we shall examine the {\it physical version process } of the first law of thermodynamics for stationary axisymmetric black holes in Einstein-Maxwell {\it dark matter}
gravity. One destroys the stationary black object by throwing
matter into the abyss of a black hole, i.e., we consider both 'ordinary' and {\it dark matter} which are swallowed by the object in question.
Our main interest will constitute the changes of the black hole parameters which enable us to find the first law of thermodynamics. Of course, one should assume that
after the considered process, the object will settle to the stationary state.

The source-free Einstein-Maxwell-{\it dark matter} equations of motion are provided by
\ben
 G_{\alpha \beta} - (T_{\alpha\beta}(F) &+&T_{\alpha\beta}(B) + \alpha~T_{\alpha\beta}(F,~B)) =0,\\
 \na_\beta \Big(F^{\alpha\beta} &+& \frac{\alpha}{2} B^{\alpha\beta} \Big)=0,\\
 \na_\beta \Big(B^{\alpha\beta} &+& \frac{\alpha}{2}F^{\alpha\beta} \Big)=0.
\een
The energy momentum tensor $T_{\alpha \beta} = - \delta S/\sqrt{-g}\delta g^{\alpha \beta}$
for the adequate fields imply
\ben
T_{\mu\nu}(F) &=&  2~F_{\mu\alpha}~F^{\alpha\nu}-\frac{1}{2} g_{\mu\nu}~F^{\alpha\beta}F_{\alpha\beta},\\
 T_{\mu\nu}(B) &=& 2~B_{\mu\alpha}B^{\alpha\nu}-\frac{1}{2}g_{\mu\nu}~B^{\alpha\beta}B_{\alpha\beta},\\
 T_{\mu\nu}(F,~B) &=& 2~F_{\mu\alpha}B^{\alpha\nu}-\frac{1}{2}g_{\mu\nu}~F^{\alpha\beta}B_{\alpha\beta}.
\een
In order to achieve the expressions for the variation of mass and angular momentum for black hole, in the first step we perform variation of the Lagrangian
given by (\ref{lagr}) with respect to the fields appearing in the model. Namely, one arrives at
\ben
 \frac{\delta \mathcal{L}}{\bf \ep} &=& \Big( G_{\mu \nu}- T_{\mu \nu}(F) +T_{\mu \nu} (B)+ \alpha~T_{\mu \nu}(F,~B) \Big)\delta g^{\mu \nu}\nonumber \\ 
 &+& 4~\Big( \na_\mu F^{\mu \nu} + \frac{\alpha}{2}~\na_\mu B^{\mu \nu} \Big)~\delta A_{\nu}
+ 4~\Big( \na_\mu B^{\mu \nu} + \frac{\alpha}{2}~\na_\mu F^{\mu \nu} \Big)~\delta B_{\nu}  + d \Theta,
 \een
 where the symplectic three-form yields
 \begin{align}
 \Theta_{\alpha \beta \ga} = \ep_{\delta \alpha \beta \ga }~\bigg[ \omega^\delta -& 4~\Big( F^{\delta \nu} + \frac{\alpha}{2}~ B^{\delta \nu} \Big)~\delta A_{\nu}
 - 4~\Big(  B^{\mu \nu} + \frac{\alpha}{2}~F^{\mu \nu} \Big)~\delta B_{\nu} \bigg],
 \end{align}
 where we set $\omega_\delta=\na^\beta \delta g_{\beta \delta} - \na _\delta \delta g_{\beta}{}{}^{\beta}$.
 
Then, having in mind the relation
 \be
 \cJ_{\beta} [\phi_a,~\Lie_\xi \phi_a] = \Theta_\beta[\phi_a,~\Lie_\xi \phi_a] - \xi_\beta~\mathcal{L},
 \ee
 where $\phi_a$ denotes the adequate fields in the theory under inspection and the Lie derivative with respect to the Killing vector $\xi_m$ stands for the variation of the
 the field in question, i.e., $\Lie_\xi \phi_a = \delta \phi_a$, one can define the Noether three-form. It is given by the expression
 \be
 \cJ_{\beta \ga \delta} [\phi_a,~\Lie_\xi \phi_a] = \ep_{\beta \ga \delta \chi}~\cJ^\chi [\phi_a,~\Lie_\xi \phi_a].
 \ee
 The explicit form of $ \cJ_{\beta \ga \delta} [\phi_a,~\Lie_\xi \phi_a]$ for Einstein-Maxwell {\it dark matter} black hole is provided by the following relation:
 \ben
  \cJ_{\beta \ga \delta} [\phi_a,~\Lie_\xi \phi_a] &=& d \Big( Q^{GR}_{\beta \ga \delta} +
  Q^{(F \alpha B)}_{\beta \ga \delta} + Q^{(B \alpha F)}_{\beta \ga \delta} \Big)\nonumber \\ 
 &+& 2~\ep_{\chi \beta \ga \delta}~\bigg(
 G^\chi {}{}_ \rho - T^\chi {}{}_\rho (F) - T^\chi {}{}_\rho (B)  
 -\alpha~T^\chi {}{}_\rho (F,~B)\bigg) \xi^\rho \\ 
 &+& 4\ep_{\chi \beta \ga \delta}~(\xi^\rho A_\rho)~\na_\mu \Big( F^{\chi \mu} + \frac{\alpha}{2}B^{\chi \mu} \Big)\nonumber 
 +4\ep_{\chi \beta \ga \delta}~(\xi^\rho B_\rho)~\na_\mu \Big(B^{\chi \mu} + \frac{\alpha}{2}F^{\chi \mu} \Big).
 \een
 Having in mind that $\cJ(\xi) = d Q(\xi) + \xi^m {\bf C}_m$ \cite{gao01}, where by ${\bf C}_m$ we have denoted a three-form built of the dynamical fields $(g_{\mu \nu},~F_{\mu \nu},~B_{\mu \nu})$, we can identify
 the quantity $ Q^{GR}_{\alpha \beta} + Q^{(F \alpha B)}_{\alpha \beta} + Q^{(B \alpha F)}_{\alpha \beta}$
 as the Noether charge for the theory under consideration. Namely,
 the Noether charge for the gravitational field is equal to $Q^{GR}_{\chi \beta} = - \ep_{\chi \beta \ga \delta} \na^\ga \xi^\delta$, while for $U(1)$-gauge fields one arrives at the following
 expressions:
 \begin{align}
 Q^{(F \alpha B)}_{\xi \beta } + Q^{(B \alpha F)}_{\xi \beta }  = - 2~\ep_{\xi \beta \ga \delta}~\Big( F^{\ga \delta} + 
\frac{\alpha}{2} B^{\ga \delta} \Big)~\xi^\rho A_\rho
 - 2~\ep_{\xi \beta \ga \delta}~\Big( B^{\ga \delta} + \frac{\alpha}{2} F^{\ga \delta} \Big)~\xi^\rho B_\rho.
 \end{align}
On the other hand, one obtains the following expression for the form ${\bf C}_m$ built of the fields appearing in the theory:
\ben \label{cabcd}
C_{\beta \ga \xi \rho} &=& 2 \ep_{\delta \beta \ga \xi} ~\Big( G^\delta {}{}_ \rho - T^\delta {}{}_\rho (F) - T^\delta {}{}_\rho (B) - \alpha~T^\delta {}{}_\rho (F,~B) \Big)\nonumber \\ 
&+&
4 \ep_{\delta \beta \ga \xi} ~\na_\mu \Big( F^{\delta \mu} + \frac{\alpha}{2} B^{\delta \mu} \Big) A_\rho +
4 \ep_{\delta \beta \ga \xi} ~\na_\mu \Big( B^{\delta \mu} + \frac{\alpha}{2} F^{\delta \mu} \Big) B_\rho.
\een
Let us notice that the condition ${\bf C}_{m} = 0$ leads to the case of source-free equations of motion for Einstein-Maxwell {\it dark matter} system. However, if we consider
non-gauge field contribution to the stress energy tensor and non-zero flux of Maxwell and {\it dark matter} fields, 
obtained by the current contribution to the action $\mathcal{L}_{cur} =  4~A^\mu j_\mu + 4~B^\mu {\tilde j}_\mu$,
 we get
\ben
G_{\mu \nu} - T_{\mu \nu}(F) &-& T_{\mu \nu}(B) - \alpha T_{\mu \nu}(F,~B) = T_{\mu \nu}^\text{m} , \\
\na_\mu \Big( F^{\mu \nu} &+& \frac{\alpha}{2} B^{\mu \nu} \Big)= j^\nu_\text{m},\\ \label{dmcur}
\na_\mu \Big( B^{\mu \nu} &+& \frac{\alpha}{2} F^{\mu \nu} \Big)= {\tilde j}^\nu_\text{dm},
\een
where $T_{\mu \nu}^\text{m}$ has the interpretation of being the non-$U(1)$-gauge field contribution to to the stress-energy tensor, while
the currents bounded with the visible and {\it dark sector} are denoted, respectively by $j^\nu_\text{m},~{\tilde j}^\nu_\text{dm}$.

In our considerations $(g_{\mu \nu},~ A_\mu,~B_\mu)$ constitute
 the source-free equations of motion of the underlying system. On the other hand,
$(\delta g_{\mu \nu},~\delta A_\mu,~\delta B_\mu)$ are the linearized perturbations 
fulfilling the linearized Einstein-Maxwell {\it dark matter} equations of motion with
the source terms given as ~$\delta T_{\mu \nu}^\text{m}$, $\delta j^\nu_\text{m}$, $\delta {\tilde j}^\nu_\text{dm}$, respectively. Bearing in mind the relation (\ref{cabcd}), one gets the following:
\ben
\delta  C_{\beta \ga \xi \rho} = \ep_{\chi \beta \ga \xi} ~\bigg[ 2~ \delta T^\chi {}{}_\rho^\text{m}
+ 4 A_\rho~\delta j^\chi_\text{m} + 4 B_\rho~\delta {\tilde j}^\chi_\text{dm} \bigg].
\een
The Killing vector field $\xi_{\alpha}$ describes also a symmetry of the background
matter field. Consequently, it provides the formula for the conserved quantities related to the Killing vector field, namely
\be
\delta H_{\xi} = - \int_{\Sigma} \xi^\beta \delta C_\beta + \int_{\Sigma}(\delta Q[\xi] - \xi \cdot \Theta).
\label{defh}
\ee
In the next step, let us 
choose $\xi^{\alpha}$ to be an asymptotic time translation $t^{\alpha}$. This fact authorizes us to identify $\delta H_t$ with 
 the variation of the Arnowitt-Deser-Misner (ADM) mass of the considered black hole. It implies
\ben \label{mm}
\delta M = - \int_\Sigma \ep_{\delta \beta \ga \zeta} ~[ 2 t^\rho ~\delta T^\delta{}{}_\rho^\text{m} + 4 t^\rho A_\rho~\delta j^\delta_\text{m} 
+ 4 t^\rho B_\rho~\delta {\tilde j}^\delta_\text{dm} 
+
\int_{\p \Sigma} \bigg( \delta Q[t^\alpha] - t^\alpha~\Theta_\alpha\bigg).
\een
On the other hand, for the Killing vector fields $\psi_\beta$, which is responsible
for the rotation in the adequate directions for black hole, $\delta H_\psi$ comprises the variation of the angular momenta for
{\it dark matter} Einstein-Maxwell black hole, given by
\ben 
\delta J = \int_\Sigma \ep_{\delta \beta \ga \zeta} ~[ 2 \phi^\rho ~\delta T^\delta{}{}_\rho^\text{m} + 4 \phi^\rho A_\rho~\delta j^\delta_\text{m}
+ 4 \phi^\rho B_\rho~\delta {\tilde j}^\delta_\text{dm} 
+
\int_{\p \Sigma} \bigg( \delta Q[\phi^\alpha] - \phi^\alpha~\Theta_\alpha \bigg).
\een
Having defined the asymptotic characteristics of the black hole, i.e., the ADM mass and angular momentum, one can proceed to the {\it physical version } of the first law of thermodynamics.
Let us assume that $( g_{\mu \nu},~F_{\mu \nu},~B_{\mu \nu})$ are solutions of the source-free Einstein-Maxwell {\it dark matter} stationary axisymmetric system. As we examine the
stationary axisymmetric black object the event horizon Killing vector field will be of the form
\be 
\xi^\alpha = t^\alpha + \Omega~\phi^\alpha.
\label{kill}
\ee
As we consider the {\it physical process } version of the first law of black hole thermodynamics, let us perturb the black object by throwing into it some ordinary and {\it dark matter}.
One supposes that the black hole will be not disturbed in course of action under consideration and settles down to the stationary state. The development will cause changes of the ADM
mass and angular momentum of the back hole. Moreover, the event horizon will be also modified.

Further, one assumes that $\Sigma_{0}$ is an asymptotically flat
hypersurface which ends on the black hole event horizon $\cH$.  As in \cite{gao01} we take into account the initial data on $\Sigma_{0}$ for the 
linearized perturbations of the fields $(\delta g_{\mu \nu},~ \delta F_{\beta \delta},~ \delta B_{\beta \delta})$ with $\delta T_{\mu \nu}^\text{m},$
$\delta j^\ga_\text{m}$ and $\delta {\tilde j}^\ga_\text{dm} $. 
We pointed out that $\delta T_{\mu \nu}^\text{m}$ envisages the perturbations connected with the non-$U(1)$-gauge field contribution to the stress-energy tensor.
Moreover, we restrict our considerations to the case when the sources $\delta T_{\mu \nu}^\text{m}$, $\delta j^\ga_\text{m}$ and 
$\delta {\tilde j}^\ga_\text{dm} $ tend to zero at infinity, as well as,
the initial data
$(\delta g_{\mu \nu},~ \delta F_{\beta \delta},~ \delta B_{\beta \delta})$ disappear in the nearby of the black hole event horizon on $\Sigma_{0}$, i.e.,
at the initial time the black object is unperturbed. Because of the fact that 
the considered perturbations vanish close to the boundary of $\Sigma_{0}$, based on the equation (\ref{defh}),
one arrives at the following:
\ben \label{hor1}
\delta M - \Omega~\delta J &=& - \int_{\Sigma_{0}} \ep_{\delta \beta \ga \zeta} ~[ 2 \xi^\rho ~\delta T^\delta{}{}_\rho^\text{m}
+ 4 \xi^\rho A_\rho~\delta j^\delta_\text{m} \nonumber\\ 
&+& 4 \xi^\rho B_\rho~\delta {\tilde j}^\delta_\text{dm} ]
= \int_{\Sigma_0} ~\bep_{\beta \ga \zeta}~\beta^\mu n_\mu = \int_\cH \beta^\mu k_\mu~\ep_{\delta \alpha \beta \ga },
\een
where by $n^\mu$ we have denoted a future directed unit normal to the hypersurface $\Sigma_0$ for black object, while
$\bep_{\alpha \beta \ga } = n^\delta ~\ep_{\delta \alpha \beta \ga }$. In the last term of the above relation, we have replaced $n^\beta$ for $k^\beta$, where
$k^\beta$ denotes the tangent vector to the null geodesic congruence of the event horizon. It is justified by the fact that 
$\beta^\delta$ given explicitly by
$\beta^\delta =
2 \xi^\rho ~\delta T^\delta{}{}_\rho^\text{m}
+ 4 \xi^\rho A_\rho~\delta j^\delta_\text{m}
+ 4 \xi^\rho B_\rho~\delta {\tilde j}^\delta_\text{dm}$ ,
is conserved and the assumption that all
the ordinary and {\it dark matter} are swallowed by the black hole.
Then the last term in the equation (\ref{hor1}) can be rewritten in the form as
\ben
\int_{\cH}\beta^\mu k_\mu~\bep_{\beta \ga \zeta}= 2 \int_\cH \delta T^\alpha{}{}_\ga^\text{m}~\xi^\ga k_\alpha~\bep_{\beta \ga \zeta} +
\Phi_{BH}^{(m)}~\delta Q^\text{(m)} 
+ \Phi_{BH}^{(dm)}~\delta Q^\text{(dm)},
\een
where we set
\ben \label{hor2}
\delta Q^\text{(dm)} = -4\int_\cH \delta {\tilde j}^\mu_\text{dm} k_\mu~\bep_{\beta \ga \zeta},&\;\;\;\;&
\delta Q^\text{(m)} = -4\int_\cH \delta {j}^\mu_\text{m} k_\mu~\bep_{\beta \ga \zeta} ,\\
\Phi_{BH}^{(m)} = - A_\beta~\xi^\beta \mid_\cH,&\;\;\;\;\;&
\Phi_{BH}^{(dm)} = - B_\alpha~\xi^\alpha \mid_\cH
\een
The final form of (\ref{hor2}) stems from the fact that for both $U(1)$-gauge fields we have the following relations:
\ben
\na_\mu (A_\beta~\xi^\beta) = \cL_\xi A_\mu + \xi^\beta F_{\beta \mu},\\ \na_\mu (B_\beta~\xi^\beta) = \cL_\xi B_\mu + \xi^\beta B_{\beta \mu},
\een
and because the Killing vector field $\xi_\alpha$ describes the symmetry of the background solution, one gets $\cL_\xi A_\mu = \cL_\xi B_\mu =0$.
Moreover, bearing in mind the Raychaudhuri equation and the fact that the shear and expansion vanish in the stationary background, one obtains that
$R_{\mu \nu}k^\mu k^\nu \mid_\cH = 0$. It implies further that for Maxwell and the {\it dark matter} gauge fields,
we have the condition $F_{\mu \beta}F_{\nu}{}{}^\beta k^\mu k^\nu \mid_\cH = 0$ and  
$B_{\mu \beta}B_{\nu}{}{}^\beta k^\mu k^\nu \mid_\cH = 0$. Consequently, we arrive at the conclusion that $F_{\beta \ga} k^\beta = 0$ and
$B_{\beta \ga} k^\beta = 0$. On the other hand, the asymmetry of the strength tensor for Maxwell and {\it dark matter} gauge fields enables us to write
$F_{\beta \ga} k^\beta \sim k_\ga$ and $B_{\beta \ga} k^\beta \sim k_\ga$. The pullbacks of $F_{\beta \ga} k^\beta $ and $B_{\beta \ga} k^\beta $ to the black hole event horizon
disappear.  Therefore the pullback of $\na_\mu \Phi_{BH}^{(m)} $ and $\na_\mu \Phi_{BH}^{(dm)}$ to the event horizon are equal to zero. It implies that on $\cH$ one has the constancy
of $\Phi_{BH}^{(m)} $ and $\Phi_{BH}^{(dm)}$.

Using the Raychaudhuri equation and the fact that the null generators of the perturbed black hole event horizon match the null generators of the event horizon of the unperturbed black object,
the following relation can be addressed \cite{wal94}
\be
 \int_\cH \delta T^\alpha{}{}_\ga^\text{m}~\xi^\ga k_\alpha~\bep_{\beta \ga \zeta}  = \kappa~\delta A,
\ee 
where $A$ is the area of the event horizon of a black hole while $\kappa$ denotes its surface gravity.

Summarizing, the {\it physical process} version of the first law for stationary axisymmetric black holes in 
Einstein-Maxwell {\it dark matter} gravity is provided by
\ben
\delta M -\Omega~\delta J - \Phi_{BH}^{(m)}~\delta Q^\text{(m)} -\Phi_{BH}^{(dm)}~\delta Q^\text{(dm)} = 2~\kappa~\delta A.
\label{1law}
\een
The above relation constitutes the main conclusion of the above derivations.
From the equation (\ref{1law}) it can be seen that the key contribution to the mass of the black hole stems from the {\it dark matter} sector coupled to the ordinary
Maxwell field. 
In the physical version of the first law the contribution of {\it dark matter} is connected with $\Phi_{BH}^{(dm)}$ and $\delta Q^\text{(dm)} $, which
constitutes the influence of $U(1)$ {\it dark sector } gauge field and the {\it dark} charge current (\ref{dmcur}). The relation (\ref{dmcur}) envisages also the influence
of the $\alpha $ coupling constant on the process in question.

As it was justified by the direct astrophysical observations \cite{mas07}-\cite{die12},
in the early Universe the {\it dark matter} formed the scaffolding 
on which the ordinary matter condensate.
Formation of black holes by condensation of the {\it dark matter} and ordinary one,
then accretion of both types of matter may play key role in growing mass of black objects to the
great extent. From the relation (\ref{1law}) it can be seen that the influence of the {\it dark matter}
on the process of early black hole formation may explain the riddle connecting explanation how such 
a giant can be formed in a relative short time after Big Bang.

The appearance of supermassive black holes at early stages of the Universe history is a 
challenge to the contemporary understanding of star and black hole formations.
So far roughly over forty quasars with redshift greater than six have been detected, each of 
them harbored by a supermassive black hole with a mass over one billion of solar masses, when
the Universe was less than one billion years old \cite{fan13}-\cite{tho16}.  However, it can be 
only a tip of the iceberg, because of the fact that black hole growth and evolution in infant Universe may 
be hidden from our contemporary view \cite{tre11}.

\section{Equilibrium state version of the first law}
This section will be devoted to the derivation of the first law of black hole dynamics in the theory under consideration, 
by choosing an arbitrary cross section of the adequate event horizon of each black object to the future
of the bifurcation sphere, in order to confront the results given by the relation (\ref{1law}).  

As it was shown in \cite{gao03} one can treat fields which were not necessarily smooth through the event horizon.
The only requirement which should be satisfied is
that the pull-back of the fields in question in the future of the bifurcation surface is smooth. For the $U(1)$ gauge fields in the considered theory with {\it dark matter} that will be the case.
In what follows we suppose that the spacetime under consideration fulfills asymptotic conditions at infinity being equipped with the Killing vector $\xi_\beta$. The Killing vector field
introduces an asymptotic symmetry \cite{iye94}, i.e., there exists a conserved quantity $H_{\xi_\beta}$ bounded with the symmetry generated by $\xi_\beta$
\be
\delta H_{\xi } = \int_{\infty} \bigg( \bdel Q(\xi ) - \xi  \cdot \Theta \bigg).
\label{qua}
\ee
Here $\bdel$ is the variation which has no effect on $\xi_{\beta}$ since the considered Killing
vector field is attended to a fixed background and it ought not to be varied in the above expression (\ref{qua}).

In the considerations we shall treat the stationary axisymmetric black object with respect to the Killing vector field (\ref{kill}). Moreover one examines asymptotic hypersurface
$\Sigma$ terminating on the part of the event horizon ${\cal H}$, to the future of the bifurcation surfaces. The inner boundary of the hypersurface $\Sigma$, ~$S_{\cH}$ will constitute the cross section
of the black hole event horizon. 
Similarly as in \cite{gao03} we shall compare variations between two adjacent states of black objects in question. 

In comparison of the two states of black holes there is a freedom 
which points can be picked up to correspond. In what follows,
we consider the case when the null vector remains normal to the hypersurface $S_{\cH} $, i.e., we make the hypersurface to be the event horizon and
$t_\beta,~\phi_\beta$ Killing vectors are the same, in the two aforementioned black hole solutions.
Just, $\delta t_\beta= \delta \phi_\beta = 0$ and the variation of the Killing vector $\xi_\beta$ will be of the form
$\delta \xi_\beta = \delta \Omega~ \phi_\beta$.

As in the previous section 
$(g_{\mu \nu}, ~F_{\beta \ga},~B_{\beta \ga})$ are solutions of Einstein-Maxwell {\it dark matter} equations of motion while their variations
$(\delta g_{\mu \nu},~\delta F_{\beta \ga},~\delta B_{\beta \ga})$ constitute perturbations fulfilling the equation of motion for the considered system.
One also requires that the pull-backs
of $F_{\beta \ga},~B_{\beta \ga}$ to the future of the bifurcation surface are smooth, but not 
necessarily smooth on it \cite{gao03}. The variations of the gauge fields fall off sufficiently rapid at infinity and these fields do not contribute to the canonical energy and canonical momenta.
In our case
\be
\delta M - \Omega~\delta J = \int_{S_{\cH}} \bigg( \bdel Q[\xi] - \xi \cdot \Theta \bigg),
\ee
where we have denote the variation $\bdel$ by the expression
\be
\bdel \int_{S_\cH} Q[\xi] = \delta \int_{S_\cH} Q[\xi] - \int_{S_\cH} Q[\delta \xi],
\ee
The Noether charge can be split into the adequate parts
\be
Q_{\beta \ga} = Q_{\beta \ga}^{(GR)} + Q_{\beta \ga}^{(F \alpha B)} + Q_{\beta \ga}^{(B \alpha F)},
\ee
where we set the following definitions for the adequate charges:
\ben
Q_{\beta \ga}^{(GR)} &=& - \ep_{\beta \ga \eta \zeta} \na^\eta \xi^\zeta, \\ \label{qfb}
Q_{\beta \ga}^{(F \alpha B)} &=& -2 \ep_{\beta \ga \eta \zeta} ~\Big( F^{\eta \zeta} + \frac{\alpha}{2} B^{\eta \zeta} \Big)~A^\mu \xi_\mu, \\ \label{qbf}
Q_{\beta \ga}^{(B \alpha F)} &=& -2 \ep_{\beta \ga \eta \zeta} ~\Big( B^{\eta \zeta} + \frac{\alpha}{2} F^{\eta \zeta} \Big)~B^\mu \xi_\mu.
\een
The arguments quoted in the previous section provide that the following is satisfied:
\ben
\int_{S_\cH} Q_{\beta \ga}^{(F \alpha B)}+ \int_{S_\cH} Q_{\beta \ga}^{(B \alpha F)}  = \Phi_{BH}^{(m)} \Big( Q^{(F)}
+\frac{\alpha}{2} Q^{(B)} \Big)
+ \Phi_{BH}^{(dm)} \Big( Q^{(B)} + \frac{\alpha}{2} Q^{(F)} \Big),
\een
where the total charges connected with the gauge fields are defined by
\be
Q^{(F)} = 2 \int_{S_\cH} \ep_{\alpha \beta \ga \delta} F^{\ga \delta}, \qquad Q^{(B)} = 2 \int_{S_\cH} \ep_{\alpha \beta \ga \delta} B^{\ga \delta}.
\ee
The variation $\bdel$ of the adequate quantities implies
\ben
\bdel \int_{S_\cH} \bigg( Q_{\delta \ga}^{(F \alpha B)} + Q_{\delta \ga}^{(B \alpha F)} \bigg)&=&
 - 4 \int_{S_\cH} \ep_{\delta \ga} 
N_\mu \xi_\nu~\Big( F^{\mu \nu} + \frac{\alpha}{2} B^{\mu \nu} \Big) 
- 4 \int_{S_\cH} \ep_{\delta \ga} 
N_\mu \xi_\nu~\Big( B^{\mu \nu} + \frac{\alpha}{2} F^{\mu \nu} \Big)\nonumber \\ 
&+& \Phi_{BH}^{(m)} \Big( \delta Q^{(F)} + \frac{\alpha}{2} \delta Q^{(B)} \Big)
+\Phi_{BH}^{(dm)} \Big( \delta Q^{(B)} + \frac{\alpha}{2} \delta Q^{(F)} \Big),
\een
where $\ep_{\beta \ga}$ is the volume element on the hypersurface $S_\cH$, while $N_{\beta}$ is the ingoing future directed null normal to $S_\cH$,
fulfilling the normalization condition of the form $N_\beta \xi^\beta = -1$. Consequently we arrive at  the following expression:
\begin{equation*}
\int_{S_\cH} \xi^{\beta} \bigg( \Theta_{\beta \delta \ga}^{(F \alpha B)}  + \Theta_{\beta \delta \ga}^{(B \alpha F)} \bigg)=
4 \int_{S_\cH} \ep_{\delta \ga} 
N_\mu \xi_\nu~\Big( F^{\mu \nu} + \frac{\alpha}{2} B^{\mu \nu} \Big)
+ 4 \int_{S_\cH} \ep_{\delta \ga} 
N_\mu \xi_\nu~\Big( B^{\mu \nu} + \frac{\alpha}{2} F^{\mu \nu} \Big). 
\end{equation*}
Further, taking into account the symplectic three-forms referred to the combinations of the $U(1)$-gauge fields and having in mind that
for each gauge field one has that
\be
F_{\mu \beta}~\xi^\beta \sim \xi_\mu, \qquad B_{\mu \beta}~\xi^\beta \sim \xi_\mu,
\ee
one can find that
\ben
\bdel \int_{S_\cH} \bigg( Q_{\delta \ga}^{(F \alpha B)} + Q_{\delta \ga}^{(B \alpha F)} \bigg)  - \int_{S_\cH} \xi^{\beta} \bigg( \Theta_{\beta \delta \ga}^{(F \alpha B)}  + 
\Theta_{\beta \delta \ga}^{(B \alpha F)} \bigg)
= \Phi_{BH}^{(m)} \delta Q^\text{(m)}\nonumber + \Phi_{BH}^\text{(dm)} \delta Q^\text{(dm)},
\een
where we denoted by $ \delta Q^\text{(m)}$ and $ \delta Q^\text{(dm)}$, respectively
\ben \label{dark}
\delta Q^{(F)} + \frac{\alpha}{2} \delta Q^{(B)}  = \delta Q^\text{(m)},\;\;\;\;\; 
 \delta Q^{(B)} + \frac{\alpha}{2} \delta Q^{(F)}  =  \delta Q^\text{(dm)}.
 \een
On the other hand, for gravitational field, one receives the following:
\be
\bdel \int_{S_\cH} Q_{\delta \ga}^{(GR)} - \xi^{\beta} ~\Theta_{\beta \delta \ga}^{(GR)} = 2\kappa~\delta A,
\ee
where $A=\int_{S_\cH} \ep_{\alpha \beta}$ is the area of the black hole event horizon.

One concludes that the equilibrium state version of the first law of thermodynamics for stationary axisymmetric black holes in
Einstein-Maxwell {\it dark matter} gravity yields
\ben
\delta M - \Omega ~\delta J - \Phi_{BH}^\text{(m)} \delta Q^\text{(m)} - \Phi_{BH}^\text{(d)} \delta Q^\text{(dm)}
= 2\kappa~\delta A.
\een
By the rigorous mathematical derivations we confirmed that the {\it equilibrium state version} of the first law of black object thermodynamics 
confirmed the results gained in the previous section.
The {\it dark matter } influence 
can be seen by the inspection of the relations (\ref{dark}), which envisage the influence of the $\alpha$ coupling constant, binding the visible and {\it dark sector},
on the adequate charges and potentials.

Hence, we can conclude that {\it dark matter} intensively influences on the mass of black holes.  As its abundance was larger in the early Universe it authorizes the key role in growing of the black object masses
at the infant phase of our Universe history.

\section{Yang-Mills with {\it dark matter} sector}
In this section we shall provide the Yang-Mills description of the Einstein gravity coupled to the {\it {\it dark matter} } sector. Among all, such a model is widely used for example in the holographic 
description of p-wave superconductors and superfluids influenced by {\it dark matter} sector \cite{rog15,rog16}.

The action of the model in question is provided by
\be
\mathcal{L} = {\bf \ep} ~\bigg( R - F_{\mu\nu}^{(a)}~F^{\mu\nu (a)} - B_{\mu\nu}^{(a)}~B^{\mu\nu (a)} - \alpha~F_{\mu\nu}^{(a)}B^{\mu\nu (a)} \bigg),
\label{lag ym}
\ee
where the strength tensor $F_{\mu\nu}^{(a)} = \na_\mu A^{(a)}_\nu - \na_\nu A^{(a)}_\mu + \ep^{abc}A^{(b)}_\mu A^{(c)}_\nu$ is used for Yang-Mills strength, while
for the {\it dark matter} $SU(2)$-gauge field we apply the analogous definition built of $B_{\mu\nu}^{(a)}$.

Variation of the Lagrangian density
\ben
 \frac{\delta \mathcal{L}}{\bf \ep} &=& \Big( G_{\mu \nu}- (T_{\mu \nu}(F) +T_{\mu \nu} (B)+ \alpha~T_{\mu \nu}(F,~B) \Big)\delta g^{\mu \nu} \\ \nonumber
 &+& 4~\Big( \na_\mu F^{\mu \nu (a)} + \frac{\alpha}{2}~\na_\mu B^{\mu \nu (a)}
 + \frac{\alpha}{2} \ep^{abc} A^{(b)}_\mu B^{\mu \nu (c)}  + \ep^{abc} A^{(b)}_\mu F^{\mu \nu (c)}
 \Big)~\delta A_{\nu} ^{(a)} \\ \nonumber
&+& 4~\Big( \na_\mu B^{\mu \nu (a)} + \frac{\alpha}{2}~\na_\mu F^{\mu \nu (a)}    + \frac{\alpha}{2} \ep^{abc} B^{(b)}_\mu F^{\mu \nu (c)}
+ \ep^{abc} B^{(b)}_\mu B^{\mu \nu (c)}    
\Big)~\delta B_{\nu}  + d \Theta^{(a)},
 \een
leads to the following equations of motion
\ben
G_{\mu \nu} - T_{\mu \nu}(F) &-& T_{\mu \nu}(B) - \alpha ~T_{\mu \nu}(F,~B) = 0, \\
\na_\mu \Big( F^{\mu \nu (a)} &+& \frac{\alpha}{2} B^{\mu \nu (a)}  \Big)
+ \frac{\alpha}{2} \ep^{abc} A^{(a)}_\mu B^{\mu \nu (a)}
+ \ep^{abc} A^{(a)}_\mu F^{\mu \nu (a)} = 0,\\
\na_\mu \Big( B^{\mu \nu (a)} &+& \frac{\alpha}{2} F^{\mu \nu (a)} \Big)
+ \frac{\alpha}{2} \ep^{abc} B^{(a)}_\mu F^{\mu \nu (a)}
+ \ep^{abc} B^{(a)}_\mu B^{\mu \nu (a)} = 0.
\een
On the other hand,  for $\Theta^{(a) \mu}$ we get the relation of the form
\ben
\Theta^{(a) \mu} = \omega^\mu - 4 \Big( F^{\mu \beta (a)} + \frac{\alpha}{2} B^{\mu \beta (a)} \Big) \delta A^{(a)}_\beta
- 
4 \Big( B^{\mu \beta (a)} + \frac{\alpha}{2} F^{\mu \beta (a)} \Big) \delta B^{(a)}_\beta.
\een

By analogy with the previous calculations we try to compute the expression
\ben
&\bdel& \int_\cH Q_{\mu \nu}^{(a)~(F\alpha B)}[\xi] + \bdel \int_\cH Q_{\mu \nu}^{(a) (B \alpha F)}[\xi]
-\int_\cH \xi^\delta \Theta^{(a)~(F\alpha B)}_{\delta \mu \nu} - \int_\cH \xi^\delta \Theta^{(a)~(B\alpha F)}_{\delta \mu \nu} = \\ \nonumber
 &-& 2 \int_{\cH} \delta \bigg[ \ep_{\mu \nu \rho \delta} \Big( F^{\rho \delta (a)} + \frac{\alpha}{2} B^{\rho \delta (a)} \Big)~A^{\beta (a)} \xi_\beta
- 2 \int_{\cH} \delta \bigg[ \ep_{\mu \nu \rho \delta} \Big( B^{\rho \delta (a)} + \frac{\alpha}{2} F^{\rho \delta (a)} \Big)~B^{\beta (a)} \xi_\beta,
\een
where $Q_{\mu \nu}^{(a)~(F\alpha B)}$ and $Q_{\mu \nu}^{(a) (B \alpha F)}$ have the same form as 
in the equations (\ref{qfb})-(\ref{qbf}), with the replacement of $U(1)$-gauge field for the $SU(2)$ one.

To commence with, let us find firstly the integrals over asymptotic hypersurface at infinity from the adequate quantities. Namely
\be
\int_\infty \Big( Q_{\mu \nu}^{(a)~(F\alpha B)}[t]  + Q_{\mu \nu}^{(a) (F\alpha B)}[t] \Big).
\ee
We remark that for the stationary axisymmetric black hole which is the solution of Einstein Yang-Mills {\it dark matter} equations of motion, one has that
$A^{(a)}_0$ and  $B^{(a)}_0$are asymptotically constant. This fact enables one to define
\be
V_{(F)} = \lim_{r \rightarrow \infty} \bigg( A^{(a)}_0~A^{(a)}_0 \bigg)^{\frac{1}{2}}, \qquad
V_{(B)} = \lim_{r \rightarrow \infty} \bigg( B^{(a)}_0~B^{(a)}_0 \bigg)^{\frac{1}{2}}.
\ee
Moreover, we can specify the notion of 'electric fields' connected with ordinary and {\it dark matter} Yang-Mills fields
\be
E^{(a)}_\beta = \sqrt{ h}~F^{\beta \mu (a)} ~n_\mu, \qquad
B^{(a)}_\beta = \sqrt{ h}~B^{\beta \mu (a)} ~n_\mu,
\ee
where $n^\mu$ is the unit normal to the spacelike hypersurface at infinity. The adequate charges measured on spacelike hypersurface at infinity can be written as
\be
Q^{\infty}_{(F)} = 4~\int_\infty \mid E^{(a)}_\beta~r^\beta \mid, \qquad Q^{\infty}_{(B)} = 4~\int_\infty \mid B^{(a)}_\beta~r^\beta \mid.
\ee
The above is sufficient to establish that
\ben
\int_\infty t^\ga ~\Theta^{(a)~(F\alpha B)}_{\ga \mu \nu} &=& \bigg( Q^{\infty}_{(F)} + \frac{\alpha}{2} Q^{\infty}_{(B)}\bigg)~\delta V_{(F)},\\
\int_\infty t^\ga ~\Theta^{(a)~(B\alpha F)}_{\ga \mu \nu} &=& \bigg( Q^{\infty}_{(B)} + \frac{\alpha}{2} Q^{\infty}_{(F)}\bigg)~\delta V_{(B)}.
\een
Because of the fact that the gravitational contribution gives us the ADM mass, as well as, defining the canonical momentum as $J = - \int_\infty Q[\phi]$,
we arrive at the following form for the first law of thermodynamics
\ben  \label{ym}
2~\kappa~\delta A &=& \delta M - \Omega~\delta J + V_{(F)} \bigg( Q^{\infty}_{(F)} + \frac{\alpha}{2} Q^{\infty}_{(B)}\bigg)
+ V_{(B)} \bigg( Q^{\infty}_{(B)} + \frac{\alpha}{2} Q^{\infty}_{(F)}\bigg) \\ 
 &-& 2 \int_{\cH} \delta \bigg[ \ep_{\mu \nu \rho \delta} \Big( F^{\rho \delta (a)} + \frac{\alpha}{2} B^{\rho \delta (a)} \Big)~A^{\beta (a)} \xi_\beta
- 2 \int_{\cH} \delta \bigg[ \ep_{\mu \nu \rho \delta} \Big( B^{\rho \delta (a)} + \frac{\alpha}{2} F^{\rho \delta (a)} \Big)~B^{\beta (a)} \xi_\beta.\nonumber
\een
The mass formula given by the equation (\ref{ym}) contains two terms proportional to the variation of the adequate duals of the considered gauge fields.
The obtained form of the first law of thermodynamics resembles the form obtained in \cite{gao03}, in the case of the ordinary Einstein $SU(2)$ Yang-Mills field.
As it was pointed out the last two terms in (\ref{ym}) cannot be evaluated in the same way as in Einstein-Maxwell theory, due to the complicity of the $SU(2)$ Lie algebra. However, some gauge conditions
were proposed in \cite{ash00, cor00,mcc13} to write them in the forms as $\Phi^{YM} = \mid \xi^\beta A_\beta{}{}^{(a)} \mid$, which is constant on the event horizon. 
The recent, mathematical treatment of this question is delivered in \cite{pra17}.

On the other hand,
the dual of the gauge strength field on the horizon will be proportional to $( \xi^\beta A_\beta{}{}^{(a)}) \sim \ast F^{\beta \ga} \ep_{\beta \ga} $. If we approve this reasoning
the last two terms in (\ref{ym}) reduce to 
$\Phi^{YM}_{(F)}~\delta Q^{YM (F)}_\cH + \Phi^{YM}_{(B)}~\delta Q^{YM (B)}_\cH $, where the charges counted on the event horizon are provided by
\be
 Q^{YM (F)}_\cH \sim - \int_\cH \ast F^{\alpha \beta (a)}, \qquad Q^{YM (B)}_\cH \sim - \int_\cH \ast B^{\alpha \beta (a)}.
 \ee
However, there is no evidence that the chosen gauge is consistent with the above attitude. It leads us to the conclusion that the equation (\ref{ym}) is the form of the first law of thermodynamics for the black object in Einstein Yang-Mills theory with {\it dark matter} sector.

\section{Binary compact objects with the influence of {\it dark matter} sector}
The other tantalizing question is connected with the analysis of the potential influence of the hidden sector on compact binary systems.
The problem is also important in the light of the recent gravitational wave detection coming from the collision of two neutron stars (GW 170817) \cite{gw5}.

In order to examine an influence of {\it dark matter} on 
binary compact objects in the General Relativity framework we shall study the helically 
symmetric spacetime \cite{blackburn, detweiler} which is considered as an analog of two oppositely charged 
particles \cite{schild}. Moreover, one considers that in the spacetime in question 
equal amounts of the ingoing and outgoing radiations propagate and consequently the 
asymptotic mass increases because of the infinite energy of the radiation field. Due to this fact, helically 
symmetric spacetimes are not asymptotically flat. However,
one expects that there exists an approximate asymptotic region \cite{detweiler}, which geometry 
is that of gravitational waves propagating on a 
Schwarzschild background and the energy carried by them is small in comparison to the mass of the binary system
under consideration.

Various aspects
of the helically symmetric spacetimes 
were studied in
\cite{bona, asada, Fus, klein, torre1, torre2, beig, bicak, yoshida, whelan}. For instance, 
circular orbits of two point particles 
in post-Minkowskian spacetimes were considered in
\cite{friedman, glenz}, while conformally flat spacelike slices of the spacetime manifold 
turned out to be  nonradiative and asymptotically flat \cite{wilson1, wilson2, baum}.

In principle, one can define a conserved Noether current and 
associated with it finite Noether
charges $Q_i$ \cite{wal93, iyerwald, iyer, sorkin, Fus} using the helical 
Killing vector field. The defined charges are independent on the
two-surface $S$, on which they are determined 
under the condition that
all kind of considered matter fields and black holes are enclosed in $S$.

In our consideration we shall consider
as an ordinary matter system composed of perfect fluid sources
described by the Lagrangian \cite{schutz, brown, iyer, carter, lee, burnett}, as well as,
Maxwell field coupled to the perfect fluid, carrying an electric current \cite{Eric}. 
On the other hand, {\it dark matter} sources will be described by the additional $U(1)$-gauge field,
coupled to the Maxwell one and to the perfect fluid, carrying {\it dark} electric current.

Our main task will be to find the first law of the binary compact objects influenced by {\it dark matter}
sector.

\subsection{Compact binary system set-up}
To commence with, let $(\mathcal{M},~g_{\mu\nu})$ be a global hyperbolic spacetime. We are 
interested in Einstein-Maxwell {\it dark matter} gravity describing black holes together with
magnetized perfect-fluid and {\it dark matter} sources.
The spacetime in question is equipped with a single Killing
vector $k^\alpha$ \cite{Fus}. In such a spacetime we shall
focus on binary systems possessing helical Killing vectors of the 
form $k^\alpha=t^\alpha+\Omega\phi^\alpha$, where $t^\alpha$ is the asymptotically 
timelike Killing vector, $\phi^\alpha$ the rotational spacelike Killing vector 
that has circular orbits with parameter length $2\pi$ \cite{Fus,Eric}. The constant 
$\Omega$ is $\Omega=\frac{2\pi}{\tau}$ where $\tau$ is a fixed period of helical integral 
curves of $k^\alpha$. The helical vector field is transverse to each Cauchy surface 
of helically symmetric spacetime however such surfaces do not admit flat 
asymptotics because of the radiation produced by the binary system in equilibrium.

The ordinary (baryonic) matter system is described by a perfect fluid characterized by 
its four-velocity $u^\alpha$ with the 
normalization condition $u^\alpha u_\alpha=-1$ and the energy-momen\-tum tensor
\begin{equation}
 T^{\alpha\beta}=\rho u^\alpha u^\beta + pq^{\alpha\beta}
\end{equation}
where $q_{\alpha\beta}=g_{\alpha\beta}+u_\alpha u_\beta$ is the projection tensor orthogonal 
to the four velocity $u^\alpha$. The energy density $\rho$ and the 
pressure $p$ of the fluid are assumed to be functions of the baryon-mass density $n$ 
and the entropy $s$ per unit baryon mass, i.e., $
 p=p(n,s),~~~\rho=\rho(n,s).$
Moreover, if we fix the energy density function $\rho=\rho(n,s)$, the following first law of thermodynamics 
and equations of motion of the fluid are provided:
\begin{align}
 d\rho(n,s)=&\frac{p+\rho}{n}dn+nTds,\\
 \nabla_\alpha(nu^\alpha)=&0,\\
 \nabla_\alpha T^{\alpha\beta}=&0,
\end{align}
where $T$ is the fluid temperature.
For the brevity of the notation, in what follows, we have introduced $h=\frac{p+\rho}{n}$.

The Lagrangian 
of the elaborated system will be given by
\begin{align}\label{lagr_ns}
 \mathcal{L}=\frac{1}{16\pi}\left\{R-F^{\mu\nu}F_{\mu\nu} -
  B^{\mu\nu}B_{\mu\nu}-\alpha~F^{\mu\nu}B_{\mu\nu}\right.
  \left.+16\pi(-\rho+A^\alpha j^1_\alpha +b B^\alpha j^2_\alpha)\right\}\sqrt{-g},
\end{align}
where
because of the fact that we are interested in {\it dark matter} influence on binary system,
we add fields describing {\it dark matter} sector coupled to perfect fluid
carrying electric and {\it dark } currents. They are denoted by $j^1_\alpha$ and $j^2_\alpha$, respectively,
$b$ stands for the additional coupling.

From now on, we shall follow the notations and conventions introduced in \cite{Fus}
in order to have a comparison with the ordinary matter binary system studies. In what follows,
we are dealing with a $1$-parameter family of 
magnetized perfect-fluid Einstein-Maxwell spacetime with {\it dark matter} sector of the following form:
\begin{align}
 \mathcal{P}(\lambda):&=[g_{\ga\beta}, u^\ga(\lambda), 
n(\lambda), s(\lambda), A_{\ga}(\lambda), B_\ga(\la),j_i^\ga(\lambda)].\nonumber\\
 j&=\{1,2\}.
\end{align}
As in \cite{Fus} the Lagrangian change will be defined as follows:
\begin{equation}
 \Delta \mathcal{P}:=\frac{d}{d\lambda}\Psi_{-\lambda}\mathcal{P}(\lambda)\mid_{\lambda=0}=(\delta+\cL_\xi)\mathcal{P}.
\end{equation}
where
the Eulerian change in each of the quantity is defined by 
$\delta \mathcal{P}:=\frac{d}{d\lambda}\mathcal{P}(\lambda)$ \cite{Fus, Eric}. But in order to 
find the change 
in the quantity, at $\lambda=0$, one needs to introduce a Lagrangian displacement $\xi^\alpha$.
Moreover, let $\Psi_\lambda$ be a diffeomorphism mapping each 
trajectory (worldline), of the initial fluid to a corresponding trajectory of the configuration $\mathcal{P}(\lambda)$. 
Therefore, the tangent $\xi^\alpha(P)$ to the path 
$\lambda\rightarrow\Psi_\lambda(P)$ can be regarded as a vector joining the fluid element in a nearby configuration.

 It is straightforward to show that the variation of the Lagrangian under our consideration implies
\begin{align}
 \frac{\delta\mathcal{L}}{\sqrt{-g}}&=-nT\Delta s+\frac{1}{\sqrt{-g}}h u_\ga\Delta(n u^\ga\sqrt{-g})
 +
 \frac{\Delta(j^\ga\sqrt{-g})}{\sqrt{-g}}(A_\ga + b~B_\ga)
-\frac{\delta g_{\ga\beta}}{16\pi}\left(G^{\ga\beta}-8\pi( T^{\ga\beta}+T^{\ga\beta}_{U(1)})\right)\nonumber\\
&-\frac{1}{4\pi}\left( \nabla_\beta(F^{\ga\beta}+\frac{\alpha}{2}B^{\ga\beta}-4\pi j_1^\ga) \right)
\delta A_\ga
-\frac{1}{4\pi}\left( \nabla_\beta(B^{\ga\beta}+\frac{\alpha}{2}F^{\ga\beta}-4b\pi j_2^\ga) \right)\delta B_\ga\nonumber\\
&-\xi_\ga\left( \nabla_\beta T^{\ga\beta}-F^{\ga\beta}j^1_\beta + A^\ga\nabla_\beta j_1^\beta\right.
\left.
+b(B^\ga\nabla_\beta j_2^\beta-B^\ga j^2_\beta) \right)+\nabla_\ga\Theta^\ga,
\end{align}
where some parts have been already computed in the previous sections and in \cite{Fus, Eric}. For the simplicity, we denote by $T^{\ga\beta}_{U(1)}$ the sum of 
the three energy-momentum tensors, that is, $T^{\ga\beta}_{U(1)}=T^{\ga\beta}(F)+T^{\ga\beta}(B)+\alpha T^{\ga\beta}(F,B)$.
The surface term $\Theta^\ga$ can be written in the form as
\begin{align}
 \Theta^\ga&
=\frac{1}{16\pi}(g^{\ga\mu}g^{\beta\delta}-g^{\ga\beta}g^{\mu \delta})\nabla_\beta\delta g_{\mu \delta}
+(\rho+p)g^{\ga\beta}\xi_\beta  + A_\beta(j_1^\ga\xi^\beta-j_1^\beta\xi^\ga) +b~B_\beta(j_2^\ga\xi^\beta-j_2^\beta\xi^\ga)
\nonumber\\
 &+\frac{1}{4\pi}(F^{\ga\beta}+\frac{\alpha}{2}B^{\ga\beta})\delta A_\beta+\frac{1}{4\pi}(B^{\ga\beta}
+\frac{\alpha}{2}F^{\ga\beta})\delta B_\beta.
\end{align}
On the other hand,
the family of Noether charges on any sphere $S$ enclosing black 
holes and neutron stars is defined as
\begin{equation}
 Q=\oint_S\, Q^{\ga\beta}dS_{\ga\beta},
\end{equation}
where $Q^{\ga\beta}=-\frac{1}{8\pi}\nabla^\ga k^\beta+k^\ga \mathfrak{B}^\beta-k^\beta\mathfrak{B}^\ga$, 
while $\mathfrak{B}^\ga$ is any family of vector 
fields satisfying the equation provided by
\begin{equation}
 \frac{1}{\sqrt{-g}}\frac{d}{d\lambda}(\mathfrak{B}^\ga\sqrt{-g})=\Theta^\ga.
\end{equation}
It can be observed that
in the case of the binary system consisting of a black hole and a neutron star,
$ \mathfrak{B}^\ga$ yields
\begin{align}
 \mathfrak{B}^\ga(\lambda)&=\frac{1}{16\pi}(g^{\ga\mu}g^{\beta\delta}-g^{\ga\beta}g^{\mu\delta})
\mid_{\lambda=0}\mathring{\nabla}_\beta g_{\mu\delta}\\
 &+\frac{1}{4\pi}(F^{\ga\beta}+\frac{\alpha}{2}B^{\ga\beta})\mid_{\lambda=0}[ A_\beta(\lambda)-c_1 A_\beta(0)]
 +\frac{1}{4\pi}(B^{\ga\beta}+\frac{\alpha}{2}F^{\ga\beta})\mid_{\lambda=0}[ B_\beta(\lambda)-c_2 B_\beta(0)]\nonumber
\end{align}
where $\mathring{\nabla}_\beta$ is the covariant derivative of the metric $g_{\mu \beta}(0)$ and $c_i,\,i=\{1,2\}$ are fixed parameters. 
One chooses $\mathfrak{B}^\ga(\lambda)$ to make $Q(\la)$ finite. If we assume that the {\it dark matter} sector behaves in a similar manner like the electromagnetic 
one, it can be shown \cite{Fus, Eric} that $Q(\la =0)$ is finite and independent of the 
sphere $S$, which is supposed to enclose the fluid and black holes associated with 
electric and {\it dark matter} charges, and respective currents. The charge $Q$ under our 
consideration is a sum of the Komar charge $Q_k$, being the gravitational 
Hamiltonian for stationary and axially symmetric spacetimes, and $Q_L$, which is an additional contribution 
related to the surface term of the Lagrangian \cite{poisson}. They are defined as follows
\begin{align}
 Q_k=-\frac{1}{8\pi}\oint_S\nabla^\alpha k^\beta dS_{\alpha\beta},\;\;\;\;\;
 Q_L=\oint_S(k^\alpha\mathfrak{B}^\beta-k^\beta\mathfrak{B}^\alpha)dS_{\alpha\beta},
\end{align}
while the surface integrals over the $i$th black hole event horizon $\mathcal{B}_i$ imply, respectively
\begin{align}
 Q_{k_i}=-\frac{1}{8\pi}\oint_{\mathcal{B}_i}\nabla^\alpha k^\beta dS_{\alpha\beta},\;\;\;\;\;
 Q_{L_i}=\oint_{\mathcal{B}_i}(k^\alpha\mathfrak{B}^\beta-k^\beta\mathfrak{B}^\alpha)dS_{\alpha\beta}.
\end{align}
In order to find the change of the Noether charge $\delta Q$, we need to compute the difference 
between the charge on the sphere $S$ and the sum of the charges on the 
black holes $\mathcal{B}_i$. It can be shown \cite{Fus, Eric} that the Komar 
charges difference associated with Lagrangian and surface 
term charges are provided by
\begin{align}
 Q_k-\sum_i Q_{k_i}&=-\frac{1}{8\pi}\int_{\partial\Sigma}\nabla^\ga k^\beta dS_{\ga\beta}\\
 &=-\frac{1}{8\pi}\int_{\Sigma}\nabla_\beta\nabla^\ga k^\beta dS_{\alpha}
 =-\frac{1}{8\pi}\int_{\Sigma}R^\ga_\beta~ k^\beta~ dS_{\ga} 
 = -\frac{1}{8\pi}\int_{\Sigma}G^\ga_\beta~ k^\beta~ dS_{\ga} 
 -\frac{1}{8\pi}\int_{\Sigma}R~ k^\ga dS_{\ga}\nonumber,\\
Q_L-\sum_i Q_{L_i}&=-\frac{1}{8\pi}\int_{\Sigma}\nabla_\beta(k^\ga
\mathfrak{B}^\beta-k^\beta\mathfrak{B}^\ga) dS_{\ga},
\end{align}
where we have denoted
$\int_{\partial\Sigma}=\oint_S-\sum_i\oint_{\mathcal{B}_i}$. 

It turns out that one can rewrite all 
the above differences,
showing that the charge $Q$ is 
of the sphere $S$ enclosing all black holes and neutron stars.
This is true because of the fact that
all the integrands appearing in the volume integrands over 
$\Sigma$ are equal to zero, in the region with no matters and currents. 

At the beginning, let us focus on the Komar charge difference
associated with with the Lagrangian (\ref{lagr_ns}). Namely one arrives at
\begin{align}
  Q_k&-\sum_i Q_{k_i}=
  -\frac{1}{16\pi}\int_\Sigma\bigg(R-F^{\mu\nu}F_{\mu\nu} -
  B^{\mu\nu}B_{\mu\nu}-\alpha F^{\mu\nu}B_{\mu\nu}
  +16\pi(-\rho+A^\ga j^1_\ga +b B^\ga j^2_\ga)\bigg)k^\mu dS_\mu \nonumber\\
  &-\int_\Sigma\bigg( T^\ga_\beta +\;^{(1)}T^{\ga}_{\beta} +\;^{(2)}T^{\ga}_{\beta} +
\alpha \;^{(3)}T^{\ga}_{\beta}\bigg)k^\beta dS_\ga
-\frac{1}{8\pi}\int_\Sigma\bigg(G^\ga_\beta-8\pi(T^\ga_\beta +T^{\ga\beta}_{U(1)})\bigg)k^\beta dS_\ga\nonumber\\
  &-\frac{1}{16\pi}\int_\Sigma\bigg(F^{\mu\nu}F_{\mu\nu} +
  B^{\mu\nu}B_{\mu\nu}+\alpha F^{\mu\nu}B_{\mu\nu}
  +16\pi(\rho-A^\ga j^1_\ga -b B^\ga j^2_\ga)\bigg)k^\mu dS_\mu
\end{align}
Some terms appearing above can be expressed
in the similar way as in \cite{Eric} up to the parts including {\it dark matter} sector. Taking into 
account the differences related to the extra field, as well as, 
the terms relating to the baryonic matter, one finally deals with the difference of Komar charges
\begin{align}
  &Q_k-\sum_i Q_{k_i}=-\int_\Sigma\mathcal{L}d^3x+\int_\Sigma(\rho+p)u^\ga u_\beta v^\beta dS_\ga
  +\int_\Sigma A_\ga(j_1^\ga k^\mu -j_1^\mu k^\ga) dS_\mu + \int_\Sigma b ~B^\ga(j_2^\ga k^\mu-j_2^\mu k^\ga)dS_\mu\nonumber \\
  &-\frac{1}{4\pi}\int_\Sigma\bigg( \cL_k A_\mu \Big(F^{\ga\mu}+\frac{\alpha}{2}B^{\ga\mu}\Big)+
  \cL_k B_\mu\Big(B^{\ga\mu}+\frac{\alpha}{2}F^{\ga\mu}\Big)\bigg)dS_\ga
  -\frac{1}{8\pi}\int_\Sigma\bigg(G^\ga_\beta-8\pi(T^\ga_\beta +T^{\ga\beta}_{U(1)})\bigg)k^\beta dS_\ga\nonumber\\
  &-\frac{1}{4\pi}\int_\Sigma\bigg[k^\beta A_\beta\bigg( \nabla_\gamma(F^{\ga\mu}+\frac{\alpha}{2}B^{\ga\mu})
-4\pi j_1^\ga \bigg) 
  +k^\beta B_\beta\bigg( \nabla_\mu (B^{\ga\mu}+\frac{\alpha}{2}F^{\ga\mu})-4b\pi j_2^\ga \bigg)
  \bigg]dS_\ga\nonumber\\
  &+\frac{1}{4\pi}\int_{\partial\Sigma}\bigg( A_\mu (F^{\ga\beta}+\frac{\alpha}{2}B^{\ga\beta})+B_\mu (B^{\ga\beta}+
  \frac{\alpha}{2}F^{\ga\beta}) \bigg)k^\mu dS_{\ga\beta}\nonumber,
\end{align}
where the last term was obtained by using the Stokes' theorem. 
On the other hand, the variation of Komar charges difference implies
\begin{align}
  \delta(Q_k&-\sum_i Q_{k_i})=-\int_\Sigma\delta\mathcal{L}d^3x+\int_\Sigma\Delta\big((\rho+p)
u^\ga u_\beta v^\beta dS_\ga\big)-\frac{1}{8\pi}\delta\int_\Sigma\bigg(G^\ga_\beta-8\pi(T^\ga_\beta +T^{\ga\beta}_{U(1)})\bigg)k^\beta dS_\ga\nonumber\\
  &+\int_\Sigma\Delta\bigg( A_\ga(j_1^\ga k^\mu-j_1^\mu k^\ga) dS_\mu \bigg) +
  \int_\Sigma\Delta\bigg( bB^\ga(j_2^\ga k^\mu -j_2^\mu k^\ga)dS_\mu \bigg)\nonumber\\
 &-\frac{1}{4\pi}\delta\int_\Sigma\bigg( \cL_k A_\mu \Big(F^{\ga\mu}+\frac{\alpha}{2}B^{\ga\mu}\Big)+
  \cL_k B_\mu \Big(B^{\ga\mu}+\frac{\alpha}{2}F^{\ga\mu}\Big)\bigg)dS_\ga\nonumber\\
  &-\frac{1}{4\pi}\delta\int_\Sigma\bigg[k^\beta A_\beta\bigg( \nabla_\gamma(F^{\ga\mu}
+\frac{\alpha}{2}B^{\ga\mu})-4\pi j_1^\ga \bigg)
  +
  k^\beta B_\beta\bigg( \nabla_\mu (B^{\ga\mu}+\frac{\alpha}{2}F^{\ga\mu})-4b\pi j_2^\ga \bigg)
  \bigg]dS_\ga\nonumber\\
  &+\frac{1}{4\pi}\delta\int_{\partial\Sigma}\bigg( A_\mu (F^{\ga\beta}+\frac{\alpha}{2}B^{\ga\beta})
+B_\mu (B^{\ga\beta}+
  \frac{\alpha}{2}F^{\ga\beta}) \bigg)k^\mu dS_{\ga\beta}.
\end{align}
In the next step,
we shall express parts with the Lagrangian change in a slightly different manner, following \cite{Eric}
, i.e., one arrives at the following:
\begin{align}
  \Delta\big((\rho+p)u^\ga u_\beta v^\beta dS_\ga\big)&=h u_\beta v^\beta\Delta(nu^\ga dS_\ga)
  +v^\beta nu^\ga\Delta(hu_\beta)dS_\ga +(\rho+p)u^\ga u^\beta\cL_k\xi^\beta dS_\ga,\\
  \Delta\Big(A_\ga(j_1^\ga k^\mu -j_1^\mu k^\ga) dS_\mu \Big)&=
  (j_1^\ga k^\mu -j_1^\mu k^\ga)\Delta A_\ga dS_\mu -k^\ga A_\ga\Delta(j_1^\mu dS_\mu)\nonumber\\
 &+\bigg(A_\ga\frac{\Delta(j_1^\ga\sqrt{-g})}{\sqrt{-g}}k^\mu +A_\ga(j_1^\mu \cL_k\xi^\beta-
 j_1^\beta\cL_k\xi^\mu )\bigg)dS_\mu\\ 
 b\Delta\Big(B_\ga(j_2^\ga k^\mu -j_2^\mu k^\ga) dS_\mu \Big)&=
  b(j_2^\ga k^\mu -j_2^\mu k^\ga)\Delta B_\ga dS_\mu -bk^\ga B_\ga\Delta(j_2^\mu dS_\mu)\nonumber\\
 &+b\bigg(B_\ga\frac{\Delta(j_2^\ga\sqrt{-g})}{\sqrt{-g}}k^\mu +B_\ga(j_2^\mu \cL_k\xi^\beta-
 j_2^\beta\cL_k\xi^\mu)\bigg)dS_\mu.
 \end{align}

Finally, the above relation together with the variation of the surface terms given by
{\small
\begin{align*}
 &\delta(Q_L-\sum_i Q_{L_i})=\oint_{\partial\Sigma}(k^\ga\Theta^\beta-k^\beta\Theta^\ga)dS_{\ga\beta}
 =\int_\Sigma\nabla_\beta(k^\ga\Theta^\beta-k^\beta\Theta^\ga)dS_\ga
 =\int_\Sigma(k^\ga\nabla_\beta\Theta^\beta  -\cL_k\Theta^\ga) dS_\ga\\
 &=
\int_\Sigma k^\ga\nabla_\beta\Theta^\beta dS_\ga
 -(\rho+p)u^\ga u_\beta\cL_k\xi^\beta dS_\ga-
  A_\beta(j_1^\ga\cL_k\xi^\beta-j_1^\beta\cL_k\xi^\ga)dS_\ga
  -bB_\beta(j_2^\ga\cL_k\xi^\beta-j_2^\beta\cL_k\xi^\ga)dS_\ga,
\end{align*}}
enables us to write down the variation of the total difference. Let us recall 
that the total charge consists of the Komar charge and the surface terms $Q=Q_k+Q_L$. 
Assuming the gauge $\delta k^\alpha=0$, as well as, $\cL_k~A_\alpha=\cL_k~B_\alpha=0$, then 
in the case when the field equations together with their perturbations and 
equations of motion are satisfied, one obtains the relation of the form
\begin{align}
 \delta(Q&-\sum_i Q_i)=\int_\Sigma\bigg( nT\Delta s\sqrt{-g}-hu_\ga
\Delta(n u^\ga\sqrt{-g})+hu_\beta v^\beta\Delta(nu^\ga dS_\ga)\bigg.
 \bigg.+v^\beta\Delta(h u_\beta)nu^\ga dS_\ga \nonumber\\
 &- (j_1^\ga k^\beta-j_1^\beta k^\ga)\Delta A_\beta-k^\beta A_\beta
\Delta(j_1^\ga dS_\ga) \bigg.
 \bigg.-b(j_2^\ga k^\beta-j_2^\beta k^\ga)\Delta B_\beta-bk^\beta B_\beta\Delta(j_2^\ga dS_\ga)   \bigg)\nonumber  \\  
 &+\frac{1}{4\pi}\delta\int_{\partial\Sigma}\bigg( A_\mu \Big(F^{\ga\beta}+\frac{\alpha}{2}B^{\ga\beta}
\Big)+B_\mu \Big(B^{\ga\beta}+
  \frac{\alpha}{2}F^{\ga\beta}\Big) \bigg)k^\mu dS_{\ga\beta}.
\end{align}
The last term in the above equation turns out to be 
\begin{equation}\label{term1}
 -\frac{1}{4\pi}\delta\sum_i \oint_{\mathcal{B}_i}\bigg( A_\mu \Big(F^{\ga\beta}
+\frac{\alpha}{2}B^{\ga\beta}\Big)+B_\mu \Big(B^{\ga\beta}+
  \frac{\alpha}{2}F^{\ga\beta}\Big) \bigg)k^\mu dS_{\ga\beta}
\end{equation}
where we have used the fact that $\int_{\partial\Sigma}=\oint_S-\sum_i\oint_{\mathcal{B}_i}$, as well as,
the relation
{\small
\begin{equation}
 \frac{1}{4\pi}\oint_S\bigg( A_\mu \Big(F^{\ga\beta}+\frac{\alpha}{2}B^{\ga\beta}\Big)+B_\mu \Big(B^{\ga\beta}+
  \frac{\alpha}{2}F^{\ga\beta}\Big) \bigg)k^\mu dS_{\ga\beta}=0.
\end{equation}
}
Some terms appearing in
the black hole charges $Q_i=Q_{K_i}+Q_{L_i}$ have been already computed in \cite{Fus, Eric}. 
Then, the Komar charge implies
$
 Q_{K_i}=\frac{1}{8\pi}\kappa_i\mathcal{A}_i,
$
where $\kappa_i$ and $\mathcal{A}_i$ are surface gravity and the area of the $i$th black hole, respectively.
Its variation is simply given as
\begin{equation}\label{term2}
 \delta Q_{K_i}=\frac{1}{8\pi}\delta\kappa_i\mathcal{A}_i+\frac{1}{8\pi}\kappa_i\delta\mathcal{A}_i.
\end{equation}

It turns out that the geometry of the spacetime, electric charge, electromagnetic and {\it dark matter} fields contribute to the surface term charge $Q_{L_i}$. The geometric part 
has been calculated in \cite{Fus} while the electromagnetic contribution is given by \cite{Eric}. Taking into account the {\it dark matter} sector,
one deals with the relation
\begin{align}\label{terms}
  \delta Q_{L_i}=&-\frac{1}{8\pi}\delta\kappa_i\mathcal{A}_i+
  \frac{1}{4\pi}\oint_{\mathcal{B}_i}\bigg(\delta(k^\mu A_\mu)\Big(F^{\ga\beta}+
\frac{\alpha}{2}B^{\ga\beta}\Big)
   +
  \delta(k^\mu  B_\mu)\Big(B^{\ga\beta}+\frac{\alpha}{2}F^{\ga\beta}\Big)\bigg)dS_{\ga\beta}.
\end{align}
Just,
the total contribution $\delta Q_i=\delta Q_{K_i}+\delta Q_{L_i}$ coming from the horizon is given finally by
\begin{align*}
 \delta Q_i&=\frac{1}{8\pi}\kappa_i\delta\mathcal{A}_i-
 \frac{1}{4\pi}\oint_{\mathcal{B}_i}\bigg(k^\mu A_\mu \delta\Big(\big(F^{\ga\beta}
+\frac{\alpha}{2}B^{\ga\beta}\big)dS_{\ga\beta}\Big)
+
  k^\mu B_\mu \delta\Big(\big(B^{\ga\beta}+\frac{\alpha}{2}F^{\ga\beta}\big)dS_{\ga\beta} \Big)\bigg).
\end{align*}
The above equation can be rewritten in the form as
\begin{align}
 \delta Q_i&=\frac{1}{8\pi}\kappa_i\delta\mathcal{A}_i+\Phi^{(m)}\bigg(\delta Q^{(m)}
+\frac{\alpha}{2}\delta Q^{(dm)} \bigg)
 +\Phi^{(dm)}\bigg(\delta Q^{(dm)}+\frac{\alpha}{2}\delta Q^{E} \bigg),
\end{align}
where the following definitions have been used:
\begin{align}
 -k^\alpha A_\alpha=\Phi^{(m)} +C^{(m)},\;\;\;\;\;
 -k^\alpha B_\alpha=\Phi^{(dm)} +C^{(dm)}.
\end{align}
In the above one introduced the analogous potential term $\Phi$ for the {\it dark matter} fields while the quantities $Q^j$ are expressed as
\begin{align}
 C^j=&-\frac{1}{4\pi Q^j}\oint_S \Phi^j C_j^{\alpha\beta}dS_{\alpha\beta},\;\;\;\;\text{where}\;\;\;j=\{m,~dm\}\nonumber\\
 C_j^{\alpha\beta}=&\{\Big(F^{\alpha\beta}+\frac{a}{2}B^{\alpha\beta}\Big),\Big(B^{\alpha\beta}+\frac{a}{2}F^{\alpha\beta}\Big)\}.
\end{align}
It all leads to the conclusion that
\begin{align}
 Q^j:=\frac{1}{4\pi}\oint_SC_j^{\ga\beta}dS_{\ga\beta}
 =\frac{1}{4\pi}\bigg(\int_{\partial\Sigma}C_j^{\ga\beta}dS_{\ga\beta}+
 \sum_i\oint_{\mathcal{B}_i} C_j^{\ga\beta}dS_{\ga\beta} \bigg).
\end{align}
Then, the total charge is provided by
\begin{align}\label{ladunek}
 Q^{\text{total}}&=\frac{1}{4\pi}\int_\Sigma\bigg(\nabla_\ga\big( F^{\ga\beta}+\frac{\alpha}{2}B^{\ga\beta} +
 B^{\ga\beta}+\frac{\alpha}{2}F^{\ga\beta}\big)dS_{\beta}  \bigg) 
 +\frac{1}{4\pi}\sum_i\oint_{\mathcal{B}_i}\big( F^{\ga\beta}+\frac{\alpha}{2}B^{\ga\beta} +
 B^{\ga\beta}+\frac{\alpha}{2}F^{\ga\beta}\big)dS_{\ga\beta}\nonumber\\
 &=\frac{1}{4\pi}\int_\Sigma\bigg(j_1^\beta+bj_2^\beta  \bigg)
 +\frac{1}{4\pi}\sum_i\oint_{\mathcal{B}_i}\big( F^{\ga\beta}+\frac{\alpha}{2}B^{\ga\beta} +
 B^{\ga\beta}+\frac{\alpha}{2}F^{\ga\beta}\big)dS_{\ga\beta}
\end{align}
where we have used the Stokes' theorem and the equations of motion for the underlying system.

The 
first law relating to changes in the thermodynamic and hydrodynamic equilibrium of matter, in the electric and {\it dark matter} currents, as well as, electromagnetic and
{\it dark matter} fields together with changes in the area and electric/{\it dark matter} charge of the horizon
implies
\begin{equation}
 \delta M = \Omega\delta J + \delta Q,
\end{equation}
where, after using the standard definitions
\begin{align}
 \bar{T}&:=\frac{T}{u^t},~~~~\bar{\mu}:=\frac{\mu}{u^t~m_B}=\frac{k-Ts}{u^t},~~~~dM_B:=n ~u^\ga~ dS_\ga,\nonumber\\
 dS&:=s~dM_B,~~~~dD_\ga:=hu_\ga~ dM_B,
\end{align}
the change of the total charge $\delta Q$ may be explicitly given by
\begin{align} \label{main}
 \delta Q&=\int_\Sigma\bigg( \bar{T}dS+\bar{\mu}\Delta dM_B
 +v^\ga\Delta dD_\ga 
\bigg.-(j_1^\ga k^\beta-j_1^\beta k^\ga)\Delta A_\beta-k^\beta A_\beta\Delta(j_1^\ga dS_\ga) \nonumber\\
 &-b(j_2^\ga k^\beta-j_2^\beta k^\ga)\Delta B_\beta-bk^\beta B_\beta\Delta(j_2^\ga dS_\ga) \bigg) \nonumber\\
 &+\frac{1}{8\pi}\sum_i\kappa_i\delta\mathcal{A}_i +\Phi^{(m)}\delta\big( Q^{(m)}+\frac{\alpha}{2}Q^{(dm)} \big)
 +\Phi^{(dm)}\delta\big( Q^{(dm)}+\frac{\alpha}{2}Q^{(m)} \big).
\end{align}
The equation (\ref{main}) constitutes the main result which shows that the influence of {\it dark matter}
sector has its large imprint on the mechanics of the compact binary objects.

\section{Conclusions}
In our paper 
the analysis is addressed to the Einstein-Maxwell gravity with {\it dark matter} sector, which was mimicked by the additional $U(1)$-gauge field coupled to the ordinary Maxwell one.
We consider the {\it physical process} version of the first law of black hole thermodynamics by destroying the stationary black object throwing into it both ordinary and {\it dark matter}.
One assumes that the black hole is not destroyed and it settles down to the stationary configuration. Then, we elaborate the {\it equilibrium state} version of the first law of mechanics by choosing 
an arbitrary cross section of the black hole event horizon to the future of the bifurcation sphere. 
We proceed further to the studies the Yang-Mills extension of the theory in question. 

Both methods reveal the fact that the key influence on the black hole masses exerts the additional $U(1)$-gauge field corresponded to the {\it dark matter} sector.
Namely, one arrives at the following relation:
$$
\delta M -\Omega~\delta J - \Phi_{BH}^{(m)}~\delta Q^\text{(m)} -\Phi_{BH}^{(dm)}~\delta Q^\text{(dm)} = 2~\kappa~\delta A.
$$
It can be easily seen that the mass of the black object is significantly larger than in the ordinary Einstein-Maxwell gravity.
In view of the fact that a non-baryonic {\it dark matter}
constitutes over twenty percent of the mass of the observed Universe, the model in question may help to understand the recent astrophysical observations.
They revealed that only a billion
years after the Big Bang, the infant Universe was lit up by bright quasars powered by supermassive black holes. The biggest of them are of twelve to seventeen billion solar masses.
{\it Dark matter} which was a scaffolding for the early Universe structure formations \cite{mas07,die12}, 
and its abundance was very high comparing to the ordinary matter,
could play the crucial role in the growth of the dark giants in the early stages of our Universe history.

Moreover, we have also studied the influence of the {\it dark matter} sector on the binary system, for example a black hole and neutron star. As it was observed
in \cite{Eric}, in general the circulation of magnetized flow is not conserved when electromagnetic fields and electric currents are present
in neutron stars. It was discussed \cite{Fus} that in the case of lack of electromagnetic fields in a helically symmetric perfect fluid spacetimes the expression 
of the first law becomes $\delta Q=0$, while for the asymptotically flat systems (like post-Newtonian one) or spatially conformally flat system one deals with 
$\delta M=\Omega\delta J$. It is a result of the conservations of the baryon mass, entropy, circulation of the flow, 
and the area of each black hole, when one models 
the late stage of binary inspiral as a sequence of equilibrium solutions. 
Due to the presence of the electromagnetic fields and electric current we cannot simplify the obtained first law,
 as in 
the mentioned examples without any assumptions concerning the flow. However, it turns out 
that applying the generalized Kelvin theorem for ideal MHD will allow us to simplify 
the first law in the given spacetimes since the generalized circulation of magnetized flow is conserved \cite{Eric}. 

In our model we have also considered analogous {\it dark matter} circulation of the flow, which 
likely is not conserved as well. Since we model the {\it dark matter} sector in the same manner as the 
electromagnetic field contribution, it seems reasonable to put similar assumptions on the { \it dark flow}
 as for the electromagnetic one. The relativistic MHD-Euler equation is enriched with 
analogous {\it dark Lorenz} force and considering Bekenstein - Oron \cite{bek, bek2} form of the 
{\it dark} four-current, one also deals with a generalized conserved circulation for {\it dark} 
magnetized flow. Repeating the arguments of \cite{Eric} for the ideal MHD flow with the 
{\it dark} counterparts we have observed that we may also write the first law as $\delta Q=0$, for 
asymptotically flat systems. The extra assumptions on the considered binary system in equilibrium are the consequence of the geometric similarity of the {\it dark matter} field, that is, 
the conservation of {\it dark matter} circulation, {\it dark flux}, and {\it dark} charge 
for a black hole - neutron star binary system.


\acknowledgments

The authors have been partially supported by the grant of the National Science Center 
$DEC-2014/15/B/ST2/00089$.




\begin{thebibliography}{99}


%
\def\cmp#1#2#3#4{\emph{#4}, \emph{ Commun. Math. Phys.} {\bf #1} (#3) #2}
\def\lmp#1#2#3#4{\emph{#4}, \emph{ Lett. Math. Phys.} {\bf #1} (#3) #2}
\def\hpa#1#2#3#4{\emph{#4}, \emph{ Hell. Phys. Acta} {\bf #1} (#3) #2}
\def\grg#1#2#3#4{\emph{#4}, \emph{ Gen. Rel. Grav.} {\bf #1} (#3) #2}
\def\pr#1#2#3#4{\emph{#4}, \emph{ Phys. Rev.} {\bf #1} (#3) #2}
\def\prl#1#2#3#4{\emph{#4}, \emph{ Phys. Rev. Lett.} {\bf #1} (#3) #2}
\def\prd#1#2#3#4{\emph{#4}, \emph{ Phys. Rev. D} {\bf #1} (#3) #2}
\def\prb#1#2#3#4{\emph{#4}, \emph{ Phys. Rev. B} {\bf #1} (#3) #2}
\def\pre#1#2#3#4{\emph{#4}, \emph{ Phys. Rev. E} {\bf #1} (#3) #2}
\def\pl#1#2#3#4{\emph{#4}, \emph{ Phys. Lett.} {\bf #1} (#3) #2}
\def\pla#1#2#3#4{\emph{#4}, \emph{ Phys. Lett. A} {\bf #1} (#3) #2}
\def\plb#1#2#3#4{\emph{#4}, \emph{ Phys. Lett. B} {\bf #1} (#3) #2}
\def\prep#1#2#3#4{\emph{#4}, \emph{ Phys. Reports} {\bf #1} (#3) #2}
\def\phys#1#2#3#4{\emph{#4}, \emph{ Physica} {\bf #1} (#3) #2}
\def\jcp#1#2#3#4{\emph{#4}, \emph{ J. Comput. Phys.} {\bf #1} (#3) #2}
\def\jmp#1#2#3#4{\emph{#4}, \emph{ J. Math. Phys.} {\bf #1} (#3) #2}
\def\jpm#1#2#3#4{\emph{#4}, \emph{ J. Phys. A: Math. Gen.} {\bf #1} (#3) #2}
\def\cpr#1#2#3#4{\emph{#4}, \emph{ Computer Phys. Rept.} {\bf #1} (#3) #2}
\def\cqg#1#2#3#4{\emph{#4}, \emph{ Class. Quant. Grav.} {\bf #1} (#3) #2}
\def\cma#1#2#3#4{\emph{#4}, \emph{ Computers Math. Applic.} {\bf #1} (#3) #2}
\def\mc#1#2#3#4{\emph{#4}, \emph{ Math. Compt.} {\bf #1} (#3) #2}
\def\apj#1#2#3#4{\emph{#4}, \emph{ Astrophys. J.} {\bf #1} (#3) #2}
\def\apjs#1#2#3#4{\emph{#4}, \emph{ Astrophys. J. Suppl.} {\bf #1} (#3) #2}
\def\acta#1#2#3#4{\emph{#4}, \emph{ Acta Astronomica} {\bf #1} (#3) #2}
\def\apl#1#2#3#4{\emph{#4}, \emph{ Ann. Physik. (Leipzig)} {\bf #1} (#3) #2}
\def\amjp#1#2#3#4{\emph{#4}, \emph{Am. J. Phys.} {\bf #1} (#3) #2}
\def\anp#1#2#3#4{\emph{#4}, \emph{ Ann. Phys.} {\bf #1} (#3) #2}
\def\sa#1#2#3#4{\emph{#4}, \emph{ Sov. Astro.} {\bf #1} (#3) #2}
\def\sia#1#2#3#4{\emph{#4}, \emph{ SIAM J. Sci. Statist. Comput.} {\bf #1} (#3) #2}
\def\aa#1#2#3#4{\emph{#4}, \emph{ Astron. Astrophys.} {\bf #1} (#3) #2}
\def\mnras#1#2#3#4{\emph{#4}, \emph{ Mon. Not. R. Astr. Soc.} {\bf #1} (#3) #2}
\def\npb#1#2#3#4{\emph{#4}, \emph{ Nucl. Phys. B} {\bf #1} (#3) #2}
\def\prsla#1#2#3#4{\emph{#4}, \emph{ Proc. R. Soc. London, Ser. A} {\bf #1} (#3) #2}
\def\jhep#1#2#3#4{\emph{#4}, \emph{ JHEP} {\bf #1} (#2) #3}
\def\nuc#1#2#3#4{\emph{#4}, \emph{ Nuovo Cimento B } {\bf #1} (#3) #2}
\def\ijmp#1#2#3#4{\emph{#4}, \emph{ Int. J. Mod. Phys. D} {\bf #1} (#3) #2}
\def\atmp#1#2#3#4{\emph{#4}, \emph{ Adv. Theor. Math. Phys.} {\bf #1} (#3) #2}
\def\ptps#1#2#3#4{\emph{#4}, \emph{ Prog. Theor. Phys. Suppl.} {\bf #1} (#3) #2}
\def\ptp#1#2#3#4{\emph{#4}, \emph{ Prog. Theor. Phys.} {\bf #1} (#3) #2}
\def\lmp#1#2#3#4{\emph{#4}, \emph{ Lett. Math. Phys.} {\bf #1} (#3) #2}
\def\cpam#1#2#3#4{\emph{#4}, \emph{ Comm. Pure Appl. Math.}  {\bf #1} (#3) #2}
\def\adv#1#2#3#4{\emph{#4}, \emph{ Adv. Phys.}  {\bf #1} (#3) #2}
\def\zh#1#2#3#4{\emph{#4}, \emph{ Zh. Eksp. Teor. Fiz.}  {\bf #1} (#3) #2}
\def\mplb#1#2#3#4{\emph{#4}, \emph{ Mod. Phys. Lett. B} {\bf #1} (#3) #2}
\def\jams#1#2#3#4{\emph{#4}, \emph{ J. Austral. Math. Soc. B} {\bf #1} (#3) #2}
\def\appa#1#2#3#4{\emph{#4}, \emph{ Acta Phys. Polonica A} {\bf #1}, (#3) #2}
\def\nat#1#2#3#4{\emph{#4}, \emph{Nature} {\bf #1}, (#3) #2}
\def\science#1#2#3#4{\emph{#4}, \emph{Science} {\bf #1}, (#3) #2}
\def\arcmp#1#2#3#4{\emph{#4}, \emph{Annual Rev. of Cond. Matter Physics} {\bf #1}, (#3) #2}
\def\zphys#1#2#3#4{\emph{#4}, \emph{Z. Phys.} {\bf #1}, (#3) #2}
\def\ncs#1#2#3#4{\emph{#4}, \emph{Nuovo Cimento Suppl.} {\bf #1}, (#3) #2}
\def\jcap#1#2#3#4{\emph{#4}, \emph{JCAP} {\bf #1}, (#3) #2}
\def\epjc#1#2#3#4{\emph{#4}, \emph{Eur. Phys. J. C} {\bf #1}, (#3) #2}
\def\fp#1#2#3#4{\emph{#4}, \emph{Found. of Phys.} {\bf #1}, (#3) #2}



%
\def\hepph#1#2{{ hep-ph }{#1} (#2)}
\def\heaph#1#2{{ astro-ph }{#1} (#2)}
\def\hepth#1#2{{ hep-th }{#1} (#2)}
\def\grqc#1#2{{ gr-qc }{#1} (#2)}
\def\ibid#1#2#3#4{\emph{#4}, {\it ibid.} {\bf #1} (#3) #2}
\def\conphy#1#2#3#4{\emph{#4}, \emph{Contemporary Physics} {\bf #1}, (#3) #2}
%
\bibitem{wal94}
R.M.Wald, {\it Quantum Field Theory in Curved Spacetime and Black Hole Thermodynamics}, 
University of Chicago Press (Chicago, 1994).
\bibitem{bar73}
J.M.Bardeen, B.Carter and S.W.Hawking, \cmp{31}{161}{1973}{The four laws of black hole mechanics}.
\bibitem{sud92}
D.Sudarsky and R.M.Wald, \prd{46}{1453}{1992}{Extrema of mass, stationarity, and staticity, and solutions to the Einstein-Yang-Mills equations}.
\bibitem{wal93}
R.M.Wald, \prd{48}{R3427}{1993}{Black hole entropy is the Noether charge}.
\bibitem{iye94}
V.Iyer and R.M.Wald, \prd{50}{846}{1994}{Some properties of the Noether charge and a proposal for dynamical black hole entropy}.
\bibitem{iye95}
V.Iyer and R.M.Wald, \prd{52}{4430}{1995}{Comparison of the Noether charge and Euclidean methods for computing the entropy of stationary black holes}.
\bibitem{iye97}
V.Iyer, \prd{55}{3411}{1997}{Lagrangian perfect fluids and black hole mechanics}.
\bibitem{jac94}
T.Jacobson, G.Kang, and R.C.Myers, \prd{49}{6587}{1994}{On black hole entropy}.
\bibitem{jac95}
T.Jacobson, G.Kang, and R.C.Myers, \prd{52}{3518}{1995}{Increase of black hole entropy in higher curvature gravity}.
\bibitem{kog98}
J.Koga and K.Maeda, \prd{58}{064020}{1998}{Equivalence of black hole thermodynamics between a generalized theory of gravity and the Einstein theory}.
\bibitem{gao03} 
S.Gao, \prd{68}{044016}{2003}{First law of black hole mechanics in Einstein-Maxwell and Einstein-Yang-Mills theories}.

\bibitem{tow01}
P.K.Townsend and M.Zamaklar, \cqg{18}{5269}{2001}{The first law of black brane mechanics}.

\bibitem{rog05a}
M.Rogatko, \prd{71}{024031}{2005}{Staticity theorem for a higher dimensional generalized Einstein-Maxwell system}.






\bibitem{gao01}
S.Gao and R.M.Wald, \prd{64}{084020}{2001}{``Physical process version'' of the first law and the generalized second law for charged and rotating black holes}.
\bibitem{bar03}
G.Barnich, \cqg{20}{3685}{2003}{Boundary charges in gauge theories: Using Stokes theorem in the bulk}.
\bibitem{bar08}
G.Barnich and G.Compere, \jmp{49}{042901}{2008}{Surface charge in gauge theories and thermodynamical in stabilities}.




\bibitem{rog02}
M.Rogatko, \cqg{19}{3821}{2002}{Physical process version of the first law of thermodynamics for black holes in Einstein-Maxwell axion-dilaton gravity}.
\bibitem{rog05}
M.Rogatko, \prd{71}{104004}{2005}{Physical process version of the first law of thermodynamics for black holes in higher dimensional gravity}.



\bibitem{cop06}
K.Copsey and G.T.Horowitz, \prd{73}{024015}{2006}{Role of dipole charges in black hole thermodynamics}.
\bibitem{elv05}
H.Elvang, R.Emparan, and P.Figueras, \jhep{02}{2005}{031}{Non-supersymmetric black rings as thermally excited supertubes}.
\bibitem{rog05br}
M.Rogatko, \prd{72}{074008}{2005}{Black rings and the physical process version of the first law of thermodynamics}, Erratum \ibid{72}{089901}{2005}{Black rings and the physical process version of the first law of thermodynamics}.


\bibitem{rog06}
M.Rogatko, \prd{73}{024022}{2006}{First law of black ring thermodynamics in higher dimensional dilaton gravity with $p+1$ strength forms}.
\bibitem{rog07a}
M.Rogatko, \prd{75}{024008}{2007}{First law of black ring thermodynamics in higher dimensional Chern-Simons gravity}.
\bibitem{rog07b}
M.Rogatko, \prd{75}{124015}{2007}{First law of black Saturn thermodynamics}.
\bibitem{rog09}
M.Rogatko, \prd{80}{044035}{2009}{First law of p-brane thermodynamics}.


\bibitem{pra17}
K.Prabhu, \cqg{34}{035011}{2017}{The first law of black hole mechanics for fields with internal gauge freedom}.
\bibitem{dut06}
K.Dutta, S.Ray, and J.Traschen, \cqg{23}{335}{2006}{Boost mass and the mechanics of accelerated black holes}.
\bibitem{app16}
M.Appels, R.Gregory, and D.Kubizniak, \prl{117}{131303}{2016}{Thermodynamics of accelerating black holes}.
\bibitem{ast17}
M.Astorino, \prd{95}{064007}{2017}{Thermodynamics of regular accelerating black holes}.

\bibitem{Fus} J.L. Friedman, K. Uryu, M. Shibata, \prd{65}{064035}{2002}{Thermodynamics of binary black holes and neutron stars}.

\bibitem{Eric} K. Uryu, E. Gourgoulhon, C. Markakis, \prd{82}{104054}{2010}{Thermodynamics of magnetized binary compact objects}.

\bibitem{blackburn} J.K. Blackburn, S. Detweiler, \prd{46.6}{2318}{1992}{Close black-hole binary systems}.

\bibitem{detweiler} S. Detweiler, \prd{50.8}{4929}{1994}{Periodic solutions of the Einstein equations for binary systems}.
\bibitem{klein}  C. Klein, \prd{70}{124026}{2004}{Binary black hole spacetimes with a helical Killing vector}.

\bibitem{torre1}  C.G. Torre, \pre{47}{073501}{2006}{Uniqueness of solutions to the helically reduced wave equation with Sommerfeld boundary conditions}. 

\bibitem{torre2} C.G. Torre, \jmp{44}{6223}{2003}{The helically-reduced wave equation as a symmetric-positive system}.

\bibitem{beig} R. Beig, J.M. Heinzle, B.G. Schmidt, \prl{98}{121102}{2007}{Helically symmetric N-particle solutions in scalar gravity}.

\bibitem{bicak} J. Bicak, B.G. Schmidt, \prd{76}{104040}{2007}{Helical symmetry in linear systems}.

\bibitem{yoshida} S. Yoshida, B.C. Bromley, J.S. Read, K. Uryu, J.L. Friedman, \cqg{23}{S599}{2006}{Models of helically symmetric binary systems}

\bibitem{whelan} J.T. Whelan, C. Beetle, W. Landry, R.H. Price, \cqg{19}{1285}{2002}{Radiation-balanced simulations for binary inspiral}.

\bibitem{bona} S. Bonazzola, E. Gourgoulhon, J.-A. Marck, \prd{56}{7740}{1997}{Relativistic formalism to compute quasiequilibrium 
configurations of nonsynchronized neutron star binaries}.

\bibitem{asada} H. Asada, \prd{57}{7292}{1998}{Formulation for the internal motion of quasi-equilibrium configurations in general relativity}.

\bibitem{bek} J. D. Bekenstein, A. Oron, \pre{62}{5594}{2000}{Conservation of circulation in magnetohydrodynamics}.

\bibitem{bek2} J. D. Bekenstein, A. Oron, \fp{31}{895}{2001}{Extended Kelvin theorem in relativistic magnetohydrodynamics}.

\bibitem{glenz} M.M. Glenz, K. Uryu, \prd{76.2}{027501}{2007}{Circular solution of two unequal mass particles in post-Minkowski approximation}.

\bibitem{wilson1} J.R. Wilson, G.J.Mathews, \prl{75}{4161}{1995}{Instabilities in close neutron star binaries}.

\bibitem{wilson2} J.R. Wilson, G.J.Mathews, P. Marronetti, \prd{54}{1317}{1996}{Relativistic numerical model for close neutron-star binaries}.

\bibitem{baum} T.W. Baumgarte, G.B. Cook, M.A. Scheel, S.L. Shapiro, S.A. Teukolsky, \prd{57}{6181}{1998}{Stability of relativistic neutron stars in binary orbit}.

\bibitem{gw1}
B.P.Abbott et al. (LIGO Scientific Collaboration and Virgo Collaboration), \prl{116}{061102}{2016}{Observation of Gravitational Waves from a Binary Black Hole Merger}.
\bibitem{gw2}
B.P.Abbott et al. (LIGO Scientific Collaboration and Virgo Collaboration), \prl{116}{241103}{2016}{GW151226: Observation of Gravitational Waves from a 22-Solar-Mass Binary Black Hole Coalescence}.
\bibitem{gw3}
B.P.Abbott et al. (LIGO Scientific Collaboration and Virgo Collaboration), \prl{118}{221101}{2017}{GW170104: Observation of a 50-Solar-Mass Binary Black Hole Coalescence at Redshift 0.2}.
\bibitem{gw4}
B.P.Abbott et al. (LIGO Scientific Collaboration and Virgo Collaboration), \prl{119}{141101}{2017}{GW170104: GW170814: A Three-Detector Observation of Gravitational Waves from a Binary Black Hole Coalescence}.
\bibitem{gw6}
B.P.Abbott et al. (LIGO Scientific Collaboration and Virgo Collaboration), \heaph{1711.05578}{2017}{\it GW170608: Observation of a 19-Solar-Mass Black Hole Coalescence}.
\bibitem{gw5}
B.P.Abbott et al. (LIGO Scientific Collaboration and Virgo Collaboration), \prl{119}{161101}{2017}{GW170817: Observation of Gravitational Waves from a Binary Neutron Star Inspiral}.

\bibitem{mas07}
R.Massey et al., \nat{445}{286}{2007}{Dark matter maps reveal cosmic scaffolding}.
\bibitem{die12}
J.Dietrich et al., \nat{487}{202}{2012}{A filament of dark matter between two cluster of galaxies}.


\bibitem{sta16}
T.K.Starkenburg, A.Helmi and L.V.Sales, \aa{587}{A24}{2016}{Dark influences}.




\bibitem{reg15}
M.Regis, J.Q.Xia, A.Cuoso, E.Branchini, N.Fornengo, and M.Viel, \prl{114}{241301}{2015}{Particle Dark Matter Searches Outside the Local Group}.
\bibitem{ali15}
Y.Ali-Haimoud, J.Chluba, and M.Kamionkowski, \prl{115}{071304}{2015}{Constrants on Dark Matter Interactions with Standard Model Particles from Cosmic Microwave Background Spectral Distortions}.
\bibitem{bra14}
J.Bramante and T.Linden, \prl{113}{191301}{2014}{Detecting {\it dark matter} with imploding pulsars in the galactic center}.
\bibitem{ful15}
J.Fuller and C.D.Ott, \mnras{450}{L71}{2015}{Dark-matter-induced collapse of neutron stars: a possible link between fast radio bursts and missing pulsar problem}.
\bibitem{lop14}
I.Lopes and J.Silk, \apj{786}{25}{2014}{A particle {\it dark matter} footprint on the first generation of stars}.
\bibitem{nak12}
A.Nakonieczna, M.Rogatko, and R.Moderski, \prd{86}{044043}{2012}{Dynamical collapse of charged scalar field in phantom gravity}.
\bibitem{nak15a}
A.Nakonieczna, M.Rogatko, and L.Nakonieczny, \jhep{11}{2015}{012}{\it Dark matter impact on gravitational collapse of an electrically charged scalar field}.

\bibitem{ger15}
A.Geringer-Sameth and M.G.Walker, \prl{115}{081101}{2015}{Indication of Gamma-Ray Emission from the Newly Discovered Dwarf Galaxy Reticulum II}.
\bibitem{bod15}
K.K.Boddy and J.Kumar, \prd{92}{023533}{2015}{Indirect detection of {\it dark matter} using MeV-range gamma-rays telescopes}.
\bibitem{til15}
K.Van Tilburg, N.Leefer, L.Bougas, and D.Budker, \prl{115}{011802}{2015}{Search for Ultralight Scalar Dark Matter with Atomic Spectroscopy}.
\bibitem{babar14}
J.P.Lees et al., \prl{113}{201801}{2014}{Search for a Dark Photon in $e^+ e^-$ Collisions at BABAR}.
\bibitem{cha17}
J.H.Chang, R.Essig, and S.D.McDermott, \jhep{01}{2017}{107}{Revisiting Supernova 1987A constraints on dark photons}.


\bibitem{vac91}
T.Vachaspati and A.Achucarro, \prd{44}{3067}{1991}{Semilocal cosmic strings}.
\bibitem{ach00}
A.Achucarro and T.Vachaspati, \prep{327}{347}{2000}{Semilocal and electroweak strings}.


\bibitem{abe04}
S.A.Abel and B.W.Schofield, \npb{685}{150}{2004}{Brane-antibrane kinetic mixing, millicharged particles and SUSY breaking}.
\bibitem{abe08}
S.A.Abel, J.Jaeckel, V.V. Khoze, and A.Ringwald, \plb{666}{66}{2008}{Illuminating the hidden sector of string theory by shining light through a magnetic field}.
\bibitem{abe08a}
S.A.Abel, M.D.Goodsell, J.Jaeckel, V.V.Khoze, and A.Ringwald, \jhep{07}{2008}{124}{Kinetic mixing of the photon with hidden U(1)s in string phenomenology}.
\bibitem{ban17}
D.Banerjee et al., \prl{118}{011802}{2017}{Search for invisible decays of sub-GeV dark photons in missing-energy events at the CERN SPS}.








\bibitem{integral}
P.Jean {\it et al.}, \aa{407}{L55}{2003}{Early SPI/INTEGRAL measurements of 511 keV line emission from the 4th quadrant of the Galaxy}.
\bibitem{atic}
J.Chang {\it et al.}, \nat{456}{362}{2008}{An excess of cosmic ray electrons at energies of 300-800 GeV}.
\bibitem{pamela}
O.Adriani {\it et al.} (PAMELA Collaboration), \nat{458}{607}{2009}{An anomalous positron abundance in cosmic rays with energies 1.5-100 Gev}.
\bibitem{muon}
G.W.Bennett {\it et al.}, \prd{73}{072003}{2006}{Final report of the E821 muon anomalous magnetic moment measurement at BNL}.

\bibitem{dav12}
H.Davoudiasl, H.S.Lee and W.J.Marciano, \prd{85}{115019}{2012}{"Dark" Z implications for parity violation, rare meson decays, and Higgs physics}.
\bibitem{dav13}
H.Davoudiasl, H.S.Lee, I.Lewis and W.J.Marciano, \prd{88}{015022}{2013}{Higgs decays as a window into the dark sector}.

\bibitem{massey15a}
D.Harvey, R.Massey, T.Kitching, A.Taylor and E.Tittley, \science{347}{1462}{2015}{The nongravitational interactions of {\it dark matter} in colliding galaxy clusters}.

\bibitem{ach16}
B.S.Acharya, S.A.R. Ellis, G.L.Kane, B.D.Nelson, and M.J.Perry, \prl{117}{181802}{2016}{Lightest visivble-sector supersymmetric particle is likely unstable}.

\bibitem{fan13}
N.Fanidakis, A.V.Maccio, C.M.Baugh, C.G.Lacey, and C.S.Frenk, \mnras{436}{315}{2013}{The most luminous quasars do not live in the most massive {\it dark matter} haloes at any redshift}.
\bibitem{mcc}
N.J.McConnell, C.P.Ma, K.Gebhardt, S.A.Wright, J.D.Murphy, T.R.Lauer, J.R.Graham, and D.O.Richstone, \nat{480}{215}{2011}{Two ten-billion-solar-mass black holes at the centers of giant elliptical galaxies}.
\bibitem{wu15}
X.B.Wu et al., \nat{518}{512}{2015}{An ultra-luminous quasar with a twelve-billion-solar-mass black hole at redshift 6.30}.
\bibitem{tho16}
J.Thomas, C.P.Ma, N.J.McConnell, J.E.Greene, J.P.Blakeslee, and R.Janish, \nat{532}{340}{2016}{A 17-billion-solar-mass black hole in a group galaxy with a diffuse core}.

\bibitem{tre11}
E.Treister, K.Schawinski, M.Volonteri, P.Natarajan, and E.Gawiser, \nat{474}{356}{2011}{Black hole growth in the early Universe is self-regulated and largely hidden from view}.














\bibitem{rog15} 
M. Rogatko, K.I. Wysoki\'nski, \jhep{03}{2015}{215}{P-wave holographic superconductor/insulator phase
transitions affected by {\it dark matter} sector}.
\bibitem{rog16}
M. Rogatko, K.I. Wysoki\'nski, \jhep{10}{2016}{152}{Condensate flow in holographic models in the presence of dark matter}.

\bibitem{ash00}
A.Ashtekar, S.Fairhurst, and B.Krishnan, \prd{62}{104025}{2000}{Isolated horizons: Hamiltonian evolution and the first law}.
\bibitem{cor00}
A.Corichi, U.Nucamendi, and D.Sudarky, \prd{62}{044046}{2000}{Einstein-Yang-Mills isolated horizons: Phase space, mechanics, hair, and conjectures}.
\bibitem{mcc13}
S.McCormick, {\it The phase space for the Einstein-Yang-Mills equations and the first law of black hole thermodynamics}, \grqc{1302.1237}{2013}.




\bibitem{schild} A. Schild, \pr{131}{2762}{1963}{Electromagnetic two-body problem}.

\bibitem{friedman} J.L.Friedman, K. Uryu, \prd{73.10}{104039}{2006}{Post-Minkowski action for point particles and a helically symmetric binary solution}.

\bibitem{iyerwald} V. Iyer, R.M. Wald, et al. \prd{52}{4430}{1995}{Comparison of the Noether charge and Euclidean methods for computing the entropy of stationary black holes}.

\bibitem{iyer} V. Iyer, \prd{55.6}{3411}{1997}{Lagrangian perfect fluids and black hole mechanics}.

\bibitem{sorkin}  R.D. Sorkin, \prsla{435}{635}{1991}{The gravitational-electromagnetic Noether operator and the second-order energy flux}. 

\bibitem{brown}  J.D. Brown, \cqg{10}{1579}{1993}{Action functionals for relativistic perfect fluids}.

\bibitem{schutz} B.F.  Schutz, R.D. Sorkin, \anp{107}{1}{1977}{Variational aspects of relativistic field theories, with application to perfect fluids}. 

\bibitem{carter}  B. Carter, \cqg{11}{2013}{1994}{Axionic vorticity variational formulation for relativistic perfect fluids}.

\bibitem{lee} J. Lee, R.M. Wald, \jmp{31}{725}{1990}{Local symmetries and constraints}

\bibitem{burnett} G. Burnett, R.M. Wald, \prsla{430}{1878}{1990}{A conserved current for perturbations of Einstein-Maxwell space-times}.

\bibitem{poisson} E. Poisson, {\it A relativist's toolkit: the mathematics of black-hole mechanics}, Cambridge University Press, 2004.







\end{thebibliography}
\end{document}